\documentstyle[11pt,aaspp4,aabib]{article}

\received{November 6 1997}
\lefthead{Thorolfsson et al.}
\righthead{Temperature Effects on Matter in Magnetic Neutron Stars}

\begin{document}
\tighten
\title{Thomas-Fermi Calculations of Atoms\\
and Matter in Magnetic Neutron Stars II:\\
Finite Temperature Effects}
\author{
A.\ Thorolfsson\altaffilmark{1,2}, \
\"{O}.\ E.\ R\"{o}gnvaldsson\altaffilmark{3}, \
J.\ Yngvason\altaffilmark{1,4}
\and 
E.\ H.\ Gudmundsson\altaffilmark{1,5}}

\altaffiltext{1}{Science Institute,
University of Iceland, Dunhaga 3, IS-107 Reykjavik, Iceland}
\altaffiltext{2}{CETP, 4 avenue de Neptune, 94107 Saint-Maur, France}
\altaffiltext{3}{Theoretical Astrophysics Center, Juliane Maries Vej 30, DK-2100 Copenhagen {\O}, Denmark}
\altaffiltext{4}{Institut f\"ur Theoretische Physik, University of Vienna, Austria}
\altaffiltext{5}{Nordita, Blegdamsvej 17, DK-2100 Copenhagen {\O}, Denmark}

\begin{abstract}
We present numerical calculations of the equation of state for dense matter 
in high magnetic fields, using a temperature dependent Thomas-Fermi theory 
with a magnetic field that takes all Landau levels into account. Free
energies for atoms and matter are also calculated as well as profiles
of the electron density as a function of distance from the atomic
nucleus for representative values of the magnetic field strength,
total matter density, and temperature. The Landau shell structure,
which is so prominent in cold dense matter in high magnetic fields, is
still clearly present at finite temperature as long as it is less than
approximately one tenth of the cyclotron energy. This structure is
reflected in an oscillatory behaviour of the equation of state and
other thermodynamic properties of dense matter and hence also in
profiles of the density and pressure as functions of depth in the
surface layers of magnetic neutron stars. These oscillations are
completely smoothed out by thermal effects at temperatures of the
order of the cyclotron energy or higher. 
\end{abstract}

%
%

\section{Introduction}
\label{sec_introduction}

The temperature of a neutron star 
exceeds $10^{11}$ K at its birth in a 
supernova core collapse. Initially the star cools very rapidly due to a 
copious emission of neutrinos (and anti-neutrinos) but once the interior 
temperature has become of the order of $10^{9}$ K the cooling becomes 
slower although it is still dominated by neutrino emission for thousands 
of years. At later stages when the temperature has fallen sufficiently the 
emission of electromagnetic radiation from the surface becomes the dominant 
cooling mechanism (see e.g.\ \cite{PrakashEtAl97}; \cite{Pethick92}; 
\cite{NomotoTsuruta87}; \cite{UmedaEtAl94}).

An important ingredient in calculations of neutron star cooling is the
thermal structure of the envelope, i.e.\ the surface layers and the outer
crust of the star. 
Although the envelope contains only a very small fraction of the star's mass, 
it controls the heat transport from the high density interior to the surface 
and thus determines the relationship between the surface temperature and the 
temperature of the stellar core, which is isothermal for sufficiently old
stars. Among other things this relationship plays a crucial role in 
relating results of theoretical cooling calculations to observations of 
radiation from neutron star surfaces (see \cite{GudmundssonEtAl83}, and
references therein). 

If the neutron star has a high magnetic field the properties of matter close 
to the surface are strongly modified by the field as compared to the zero 
field case, thus changing both density and temperature profiles in the envelope 
(see e.g.\ \cite{RFGPY} (hereafter Paper I), and references 
therein). 
The field may also have a considerable effect on the spectrum of 
radiation emitted from the stellar surface as well as its polarization 
and angular distribution (see e.g.\ \cite{PavlovEtAl95}). 

In order to investigate the thermal structure of neutron star 
envelopes one needs as physical input both the opacity as well as the equation 
of state of matter for the relevant temperatures and densities. The case of 
iron envelopes in zero magnetic field was investigated in detail by 
\cite*{GudmundssonEtAl83}, and more recently
the thermal structure of envelopes with accreted 
layers of lighter elements has been investigated by
\cite*{ChabrierEtAl97}
and \cite*{PotekhinEtAl97}. 
Several authors have also included the effects of strong magnetic fields 
to varying degrees of accuracy (e.g.\ \cite{Hernquist85}; \cite{VanRiper88}; 
Schaaf 1990a,b\nocite{Schaaf90a}\nocite{Schaaf90b};
\cite{Page95}). The paper by \cite*{YakovlevKaminker94} gives 
a comprehensive review of the physical properties of neutron star envelopes 
with magnetic fields, including a general discussion of magnetic opacities. 
More recently \cite*{Potekhin96} and \cite*{PotekhinYakovlev96} have
published practical expressions for electric and thermal 
conductivities along magnetic fields under conditions occurring in neutron 
stars, and \cite*{LaiSalpeter97} have investigated the effects of
different phases of hydrogen on the properties of the atmospheres and
surface emission of strongly magnetized neutron stars.
   
In this paper we extend our previous work on the zero temperature equation 
of state of dense matter in high magnetic fields (Paper I) by including 
the effects of temperature for all Landau levels. As before we use the pure
Thomas-Fermi (TF) method for matter in a magnetic field, where by pure we mean
that correction terms such as the exchange (Dirac) correction and the von 
Weizs{\"{a}}cker gradient correction are not included in the calculations. 
For the problem under discussion this is in fact a good approximation as 
emphasized by \cite*{FushikiEtAl92}, who present a detailed treatment of 
magnetic TF theory and give extensive references to earlier work (see also 
Paper I and \cite{LSY94}).  

The paper is organized as follows.
In \S 2 we first give a short general overview of TF theory at arbitrary 
temperature and then specialize the discussion to hot matter in high magnetic
fields with all Landau levels taken into account. There we also discuss 
the range of validity of the theory and its scaling properties. 
Our numerical methods are described in \S 3, and in \S 4 we present the results 
of the calculations for the equation of state and the sound velocity of hot
neutron star matter as well as isothermal density profiles of surface layers
of neutron stars. As in Paper I we have chosen iron, with $Z = 26$ and 
$A = 56$, as our reference element in the numerical calculations, and the range
of magnetic field strengths covered is from $10^{10}$ to $10^{13}$ G.
Because of the scaling properties of TF theory, discussed in \S 2.5, 
this is not a serious restriction. 
The results presented in \S 4 are mainly in graphical
form but for the convenience of potential users a tabulated version can be 
found at the website http://www.raunvis.hi.is/\~{}ath/TFBT. 

Some of the bulk properties of hot dense matter in a strong magnetic field 
have previously been investigated by other authors. 
\cite*{GadiyakEtAl81} used a 
modified TF theory including more than one Landau level to calculate an 
equation of state for iron in a magnetic field of order $10^{12}$ G.
Using the strong field version of the TF-method
\cite*{ConstantinescuMoruzzi78} studied the case of hot iron in bulk in
the limit when the field
is so strong that only the lowest Landau level is populated. A similar
investigation was performed by \cite*{AbrahamsShapiro91} who also constructed
a model including exchange and gradient correction to the equation of state
of iron in the strong field limit.    

%
%
\def\greaterthansquiggle{\raise.3ex\hbox{$>$\kern-.75em\lower1ex\hbox{$\sim$}}}
\def\lessthansquiggle{\raise.3ex\hbox{$<$\kern-.75em\lower1ex\hbox{$\sim$}}}
\newcommand{\beq}{\begin{equation}}
\newcommand{\eeq}{\end{equation}}
\newcommand{\beqa}{\begin{eqnarray}}
\newcommand{\eeqa}{\end{eqnarray}}
\newcommand{\beqan}{\begin{eqnarray*}}
\newcommand{\eeqan}{\end{eqnarray*}}
\newcommand{\ba}{\begin{array}}
\newcommand{\ea}{\end{array}}
\newcommand{\no}{\nonumber}
\newcommand{\bob}{\hspace{0.2em}\rule{0.5em}{0.06em}\rule{0.06em}{0.5em}\hspace{0.2em}}
\newcommand{\grts}{\greaterthansquiggle}
\newcommand{\lets}{\lessthansquiggle}
\def\dddot{\raisebox{1.2ex}{$\textstyle .\hspace{-.12ex}.\hspace{-.12ex}.$}\hspace{-1.5ex}}
\def\Dddot{\raisebox{1.8ex}{$\textstyle .\hspace{-.12ex}.\hspace{-.12ex}.$}\hspace{-1.8ex}}
\newcommand{\Un}{\underline}
\newcommand{\ol}{\overline}
\newcommand{\ra}{\rightarrow}
\newcommand{\Ra}{\Rightarrow}
\newcommand{\ve}{\varepsilon}
\newcommand{\vp}{\varphi}
\newcommand{\vt}{\vartheta}
\newcommand{\dg}{\dagger}
\newcommand{\wt}{\widetilde}
\newcommand{\wh}{\widehat}
\newcommand{\br}{\breve}
\newcommand{\A}{{\cal A}}
\newcommand{\B}{{\cal B}}
\newcommand{\C}{{\cal C}}
\newcommand{\D}{{\cal D}}
\newcommand{\E}{{\cal E}}
\newcommand{\F}{{\cal F}}
\newcommand{\G}{{\cal G}}
\newcommand{\Ha}{{\cal H}}
\newcommand{\K}{{\cal K}}
\newcommand{\cL}{{\cal L}}
\newcommand{\M}{{\cal M}}
\newcommand{\N}{{\cal N}}
\newcommand{\cO}{{\cal O}}
\newcommand{\cP}{{\cal P}}
\newcommand{\Q}{{\cal Q}}
\newcommand{\R}{{\cal R}}
\newcommand{\cS}{{\cal S}}
\newcommand{\T}{{\cal T}}
\newcommand{\U}{{\cal U}}
\newcommand{\V}{{\cal V}}
\newcommand{\W}{{\cal W}}
\newcommand{\X}{{\cal X}}
\newcommand{\Y}{{\cal Y}}
\newcommand{\Z}{{\cal Z}}
\newcommand{\st}{\stackrel}
\newcommand{\dfrac}{\displaystyle \frac}
\newcommand{\dint}{\displaystyle \int}
\newcommand{\dsum}{\displaystyle \sum}
\newcommand{\dprod}{\displaystyle \prod}
\newcommand{\dmax}{\displaystyle \max}
\newcommand{\dmin}{\displaystyle \min}
\newcommand{\dlim}{\displaystyle \lim}
\def\QED{\\ {\hspace*{\fill}{\vrule height 1.8ex width 1.8ex }\quad}
    \vskip 0pt plus20pt}
\newcommand{\hy}{${\cal H}\! \! \! \! \circ $}
\newcommand{\h}[2]{#1\dotfill\ #2\\}
\newcommand{\tab}[3]{\parbox{2cm}{#1} #2 \dotfill\ #3\\}
\def\nz{\ifmmode {I\hskip -3pt N} \else {\hbox {$I\hskip -3pt N$}}\fi}
\def\zz{\ifmmode {Z\hskip -4.8pt Z} \else
       {\hbox {$Z\hskip -4.8pt Z$}}\fi}
\def\qz{\ifmmode {Q\hskip -5.0pt\vrule height6.0pt depth 0pt
       \hskip 6pt} \else {\hbox
       {$Q\hskip -5.0pt\vrule height6.0pt depth 0pt\hskip 6pt$}}\fi}
\def\rz{\ifmmode {I\hskip -3pt R} \else {\hbox {$I\hskip -3pt R$}}\fi}
\def\cz{\ifmmode {C\hskip -4.8pt\vrule height5.8pt\hskip 6.3pt} \else
       {\hbox {$C\hskip -4.8pt\vrule height5.8pt\hskip 6.3pt$}}\fi}
\newtheorem{theorem}{Theorem}
\newtheorem{definition}{Definition}
\newtheorem{lemma}{Lemma}
\newtheorem{cor}{Corollary}
\newtheorem{prp}{Proposition}
\def\lint{\int\limits}

\section{The Thomas-Fermi Method}
\subsection{The General TF Equation}
A quick road to TF theory at arbitrary temperature starts with the basic
thermodynamical relationship between the particle density $n$ and the
pressure $P_{\rm el}$ for noninteracting electrons
\beq
n = P'_{\rm el}(\mu)\label{density}
\eeq
where the dash denotes differentiation with respect to the chemical potential
$\mu$. For electrons that interact with each
other and fixed nuclei by Coulomb forces the electron density depends on the 
position ${\bf r}$, and the electrostatic potential generated by $K$ nuclei 
with 
charges $eZ_i$ at 
positions ${\bf R}_i$ together with the electrons is 
\beq
\Phi_n({\bf r}) = \sum_{i=1}^K \frac{eZ_i}{|{\bf r} - {\bf R}_i|} -
e \int \frac{n({\bf r}')}{|{\bf r} - {\bf r}'|} d^3 {\bf r}'\label{pot}.
\eeq
In the TF approximation it is assumed that the relation (\ref{density}) holds 
locally,
but the 
total electrochemical potential $\mu_{\rm TF}\equiv
\mu-e\Phi_n({\bf r})$ of the electrons (called
$\mu_{\rm tot}$ in Paper I)
is independent of position. 
The TF equation for the density $n({\bf r})$ thus becomes 
\beq
n({\bf r}) = P'_{\rm el}(e\Phi_n({\bf r})+\mu_{\rm TF} )\label{tfeq1}.
\eeq
The constant $\mu_{\rm TF}$ depends on the total number of electrons and the 
boundary conditions.

The nonlinear integral equation (\ref{tfeq1}) for $n$ may be combined with 
Poisson's
equation
\beq
\nabla^2 \Phi_n({\bf r}) = 4 \pi e \left[n({\bf r}) - \sum_{i=1}^K
Z_i \delta({\bf r} - {\bf R}_i) \right]\label{poisson}
\eeq
to obtain the TF differential equation for 
$\Psi \equiv e\Phi_n+\mu_{\rm TF}$ :
\beq
\nabla^2 \Psi({\bf r}) = 4 \pi e^2 P'_{\rm el}(\Psi({\bf r}))\label{tfeq2}
\eeq
for ${\bf r} \notin \{{\bf R}_1, \ldots,{\bf R}_K\}$. To take care of the
Coulomb singularities at the nuclear positions, (5) has to be
supplemented by the boundary conditions
\beq
\lim_{{\bf r} \ra {\bf R}_i} |{\bf r} - {\bf R}_i| \Psi({\bf r}) =
e^2 Z_i.\label{boundary1}
\eeq
Moreover, the whole system is confined within a bounded region $\Omega$,
and the total number of electrons within $\Omega$, $N$, is fixed.
Integrating (\ref{poisson}) over $\Omega$ we obtain by the divergence theorem
\beq
\int_{\partial \Omega} (\nabla \Psi) \cdot d{\bf S} = 4 \pi e^2
\left(N - \sum_{i=1}^K Z_i \right).\label{boundary2}
\eeq

In the spherically symmetric, neutral case considered in \S 2.4,
$\Omega$ is a sphere with radius $r_0$ containing a single nucleus
of charge $Z = N$ at the centre. Hence, the spherically symmetric $\Psi$
satisfies
\beq
\left.\frac{d}{dr} \Psi(r) \right|_{r=r_0} = 0\label{boundary3}
\eeq
in this case.

Once (\ref{tfeq2}), (\ref{boundary1}) and (\ref{boundary2}) (or 
(\ref{boundary3}))
have been solved the electronic density
may be obtained from (\ref{tfeq1}), i.e.
\beq
n({\bf r}) = P'_{\rm el}(\Psi({\bf r})).
\eeq
Likewise, the local pressure in the electron gas is
\beq
P_{\rm loc}({\bf r}) = P_{\rm el}(\Psi({\bf r})).
\eeq

\subsection{TF Theory as a Minimization Problem}
By Legendre transformation one may pass from the pressure $P_{\rm el}$
to the free energy density
\beq
f_{\rm el}(n) = \sup_\mu \{\mu n - P_{\rm el}(\mu)\}\label{free}.
\eeq
The derivative $f'_{\rm el} = \partial f_{\rm el}/\partial n$ is the
inverse of $P'_{\rm el}$, so (\ref{density}) is equivalent to
\beq
\mu= f'_{\rm el}(n)
\eeq
and the TF equation (2) may be written as
\beq
f'_{\rm el}(n({\bf r})) - e\Phi_n({\bf r})=\mu_{\rm TF}.\label{tfeq5}
\eeq
This again states that the total electrochemical potential is independent of 
the position.
Equation (\ref{tfeq5}) is 
the Euler-Lagrange equation of a minimization problem: Define
the free energy functional of the interacting electron gas as
\beq
\F[n] =
\int_\Omega f_{\rm el}(n({\bf r}))d^3 {\bf r} -
\sum_{i=1}^K e^2 Z_i \int_\Omega \frac{n({\bf r})}{|{\bf r} - {\bf R}_i|}
d^3 {\bf r} + {e^2\over 2} \int_\Omega \int_\Omega
\frac{n({\bf r}) n({\bf r}')}{|{\bf r} - {\bf r}'|} d^3 {\bf r}
d^3 {\bf r}'.\label{functional}
\eeq
Then (\ref{tfeq5}) is the variational equation for the minimization of 
(\ref{functional}) under
the subsidiary condition
\beq
\int n({\bf r}) d^3 {\bf r} = N = \mbox{constant.}
\eeq    
The $\mu_{\rm TF}$ in (\ref{tfeq5}) is the Lagrange multiplier  corresponding 
to the
subsidiary condition. It is given by
$$
\mu_{\rm TF} = \frac{\partial F(N)}{\partial N}
$$
where $F(N) = \F[n_{\rm TF}]$ is the free energy at the minimizing
density $n_{\rm TF}$, i.e., the solution to the TF equation (\ref{tfeq5}).

\subsection{The Electron Gas in a Magnetic Field at Nonzero Temperature}
We shall now specialize the general scheme described above to the case of 
non-relativistic
electrons in a strong, homogeneous magnetic field of strength $B$. For
the rest of this section and in \S 3, we
find it convenient to employ atomic units where Planck's constant 
$\hbar$,
Boltzmann's constant $k$, the electron mass $m_e$ and the elementary
charge $e$ are all taken to be 1 and dimensionless. 
This means that magnetic fields are measured in the natural unit
$$
B_0 = \frac{m_e^2 e^3 c}{\hbar^3} = 2.35 \times 10^9\mbox{ G},
$$
lengths are measured in terms of the Bohr radius
$$
a_0 = \frac{\hbar^2}{m_e e^2} = 0.529 \times 10^{-8}\mbox{ cm}
$$
and energies in the unit
$$
\frac{e^2}{a_0} = 27.2 \mbox{ eV}.
$$
The corresponding temperature unit is 
$$\frac{e^2}{a_0k} = 3.16\times 10^5\mbox{ K}$$ and pressure is measured in
$$\frac{e^2}{a_0^4} = 8.2\times 10^{26}\mbox{ dyn/${\rm cm}^2$}.$$
(When presenting our results in \S 4 we shall convert back to
conventional units.)

The motion of the electrons perpendicular to the magnetic field is
quantized into Landau levels with energy $\nu B$, $\nu = 0,1,2,\ldots$.
The degeneracy of the levels, per unit area, is $B/2\pi$ for $\nu = 0$ but
twice as high for $\nu > 0$ due to the electron spin. Along the field
the motion is one-dimensional with the density of states
$D(\ve) = \ve^{-1/2}/(2^{1/2} \pi)$, where $\ve$ is the energy of the
translational motion.

\begin{figure}
\plotone{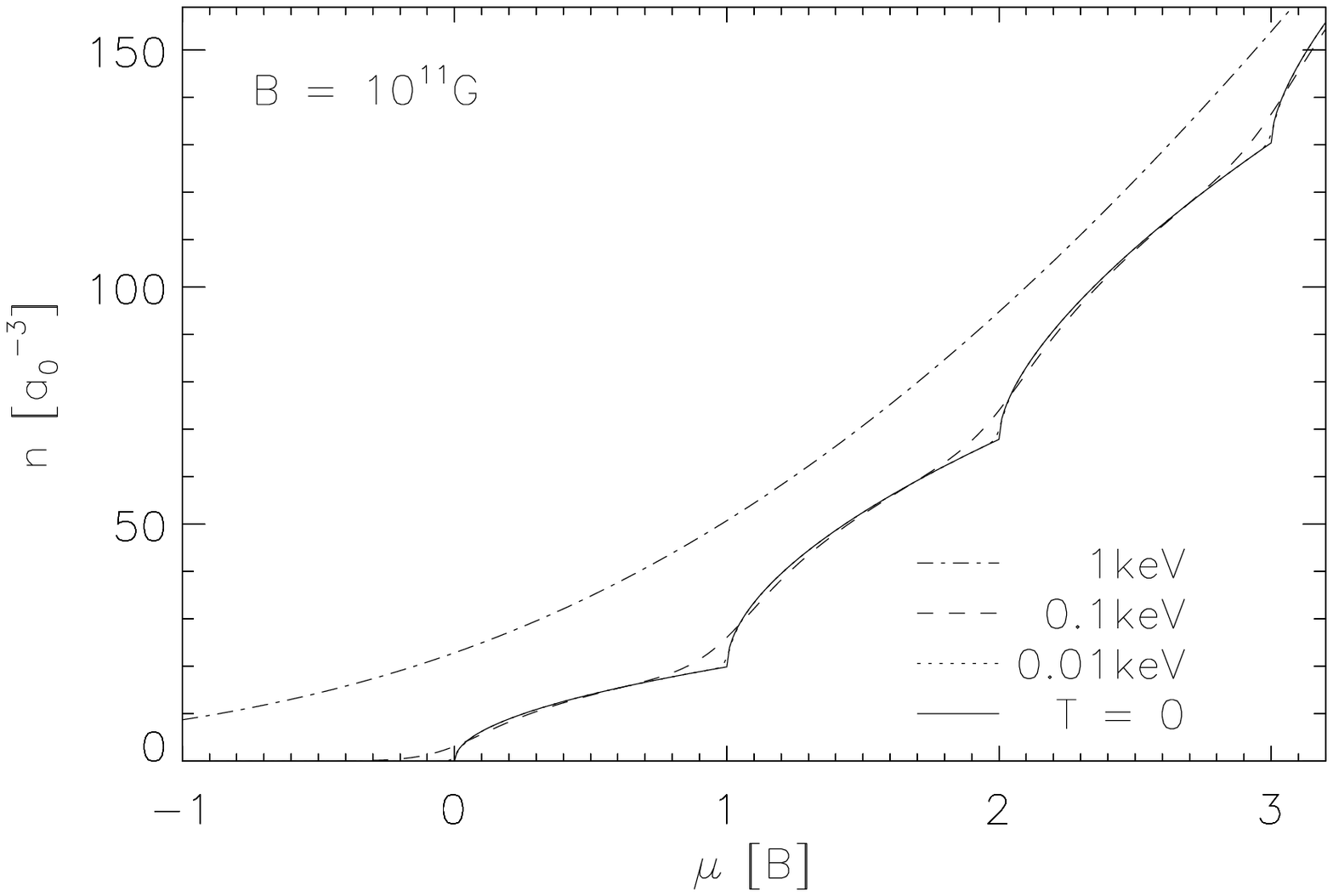}
\caption{The electron number density as a function of the chemical
potential for a uniform gas of free electrons.
The curves are drawn for $B=10^{11}$ G and three temperatures between
0.01 keV and 1 keV in addition to the zero temperature case.
Note that the
Landau band structure which is still present at $T = 0.01$ keV has almost
disappeared at $T = 0.1$ keV and is not visible at $T = 1$ keV. The quantities
on the axes are dimensionless: $n$ is written in units of $a_{o}^{-3}$ and
$\mu$ in units of the cyclotron energy (equal to $B$ in dimensionless
units).
\label{Fig_nmu}}
\end{figure}

It follows that the particle density at temperature $T$ and chemical
potential $\mu$ is given by
\beqa
n=P'_{\rm el}(\mu;T,B) &=& \frac{B}{2\pi} \frac{1}{2^{1/2}\pi}
\left[ \int_0^\infty \frac{\ve^{-1/2}}{e^{(\ve - \mu)/T}+1} d\ve 
+ 2 \sum_{\nu = 1}^\infty \int_0^\infty 
\frac{\ve^{-1/2}}{e^{(\ve + \nu B - \mu)/T} +1} \right] \no \\
&=& \frac{B T^{1/2}}{2^{3/2}\pi^2} \left[ I_{-1/2} \left(\frac{\mu}{T}
\right) + 2 \sum_{\nu =1}^\infty I_{-1/2} \left( \frac{\mu - \nu B}{T}
\right)\right]\label{pdash}
\eeqa
(see figure \ref{Fig_nmu})
where the Fermi-Dirac integral for $k > -1$ is
$$
I_k(x) = \int_0^\infty \frac{y^k}{e^{y-x} + 1} dy.
$$
Using the relation $\frac{d}{dx} I_k(x) = k I_{k-1}(x)$ we obtain 
the pressure
\beq
P_{\rm el}(\mu;T,B)
= \frac{B T^{3/2}}{2^{1/2}\pi^2} \left[ I_{1/2} \left(\frac{\mu}{T}
\right) + 2 \sum_{\nu =1}^\infty I_{1/2} \left( \frac{\mu - \nu B}{T}
\right)\right].\label{p}
\eeq
It is evident that $P_{\rm el}$ and $P'_{\rm el}$ depend in a nontrivial
way only on the ratios
\beq
\eta := \frac{\mu}{T} \qquad \mbox{and} \qquad \xi := \frac{B}{T}.
\eeq
Moreover, if $\xi \ll 1$ (weak fields and/or high $T$) and $\eta$
arbitrary, or $\xi \ll |\eta|$ (high or low densities) one may use the
approximation
\beq
\xi \sum_{\nu = 1}^\infty I_k(\eta - \nu \xi) \approx \int_0^\infty
I_k (\eta - \xi)d\xi = \frac{1}{k+1} I_{k+1}(\eta).\label{approx}
\eeq
Hence, in these limiting cases
\beq
P'_{\rm el}(\mu;T) \approx \frac{2^{1/2} T^{3/2}}{\pi^2} I_{1/2}
\left( \frac{\mu}{T}\right)\label{limit1}
\eeq
and
\beq
P_{\rm el}(\mu;T) \approx \frac{2^{3/2} T^{5/2}}{3\pi^2} I_{3/2}
\left( \frac{\mu}{T}\right),\label{limit2}
\eeq
are independent of $B$. At low densities, $\eta\to-\infty$, one may 
use the further approximation $I_{k+1}(\eta)\approx (k+1)!e^\eta$, 
and (\ref{limit1}) and (\ref{limit2}) reduce to the equation for a 
classical, ideal gas,
\beq
P_{\rm el}=nT.
\label{idealgas}
\eeq

\subsection{TF Theory in the Spherical Approximation}
In the spherical approximation to TF theory of bulk matter each
Wigner-Seitz cell is approximated by a sphere with the nucleus at the center.
Thus, one considers (\ref{tfeq2}), (\ref{boundary1}) and (\ref{boundary2}) 
with 
$\Omega$ a sphere of radius
$r_0$ and a single nucleus with charge $Z$ at ${\bf R}_1 = 0$. We
consider only the neutral case, $N = Z$. Due to the spherical symmetry
we may assume that $\Psi$ depends only on the radial coordinate
$r = |{\bf r}|$. To eliminate the nuclear charge $Z$ we define
\beq
x = Z^{1/3} r,
\eeq
\beq
\beta = \frac{B}{Z^{4/3}}, \qquad \tau = \frac{T}{Z^{4/3}},
\eeq
and it is convenient to replace $\Psi$ by
\beq
\Theta(x) = \frac{r \Psi(r)}{Z} = \frac{x \Psi(Z^{-1/3} x)}{Z^{4/3}}.
\eeq
(Note that the dimensionless radial coordinate $x$ is not the same as
the one used in Paper I whereas the dimensionless potential $\Theta$
is the same.)

The TF equation (\ref{tfeq2}) written in terms of these variables becomes
\beq
\frac{d^2\Theta(x)}{dx^2} = \frac{2^{1/2}}{\pi}  \beta \tau^{1/2}x
\left[I_{-1/2} \left( \frac{\Theta(x)}{x \tau}\right) + 
2 \sum_{\nu =1}^\infty I_{-1/2} \left( \frac{\Theta(x)}{x\tau} -
\frac{\nu \beta}{\tau} \right) \right].\label{tfeq3}
\eeq
It has two parameters, $\beta$ and $\tau$.

The approximation (\ref{limit1}) corresponds to
\beq
\frac{d^2 \Theta(x)}{dx^2} = \frac{2^{5/2}}{\pi}  \tau^{3/2}x
I_{1/2} \left( \frac{\Theta(x)}{x\tau} \right).
\eeq
It can be applied when either $\beta \ll \tau$ (high temperature and/or
weak fields), or $\beta \ll \Theta/x$ (high density close to the
nucleus).

The boundary condition (\ref{boundary1}) means that
\beq
\Theta(0) = 1\label{boundary4}
\eeq
and the boundary condition (\ref{boundary3}) is
\beq
\Theta'(x_0) = \frac{\Theta(x_0)}{x_0}\label{boundary5}
\eeq
where $x_0 = Z^{1/3} r_0$, and the dash denotes differentiation with
respect to $x$.

The pressure $P_{{\rm TF}}$ of bulk matter in the spherical TF approximation 
can be
shown to be equal to the local electronic pressure at the boundary
(\cite{FushikiEtAl89}):
\beq
P_{{\rm TF}}= P_{\rm loc}(r_0) = P_{\rm el} \left(
\frac{Z^{4/3} \Theta(x_0)}{x_0}\right)\label{tfeq4}
\eeq
and the matter density $\rho$ is simply given by:
\beq
\rho = \frac{A m_{\rm n}}{(4\pi/3) r_0^3}
     = \frac{AZm_{\rm n}}{(4\pi/3) x_0^3}\label{rho}
\eeq
where $A$ is the mass number and $m_{\rm n}$ is the nucleon mass.
Combining (\ref{tfeq4}) and (\ref{rho}) gives rise to
the TF equation of state,
\beq
P_{{\rm TF}} = P_{{\rm TF}}(\rho;T,B).\label{ptf}
\eeq

\begin{figure}
\plotone{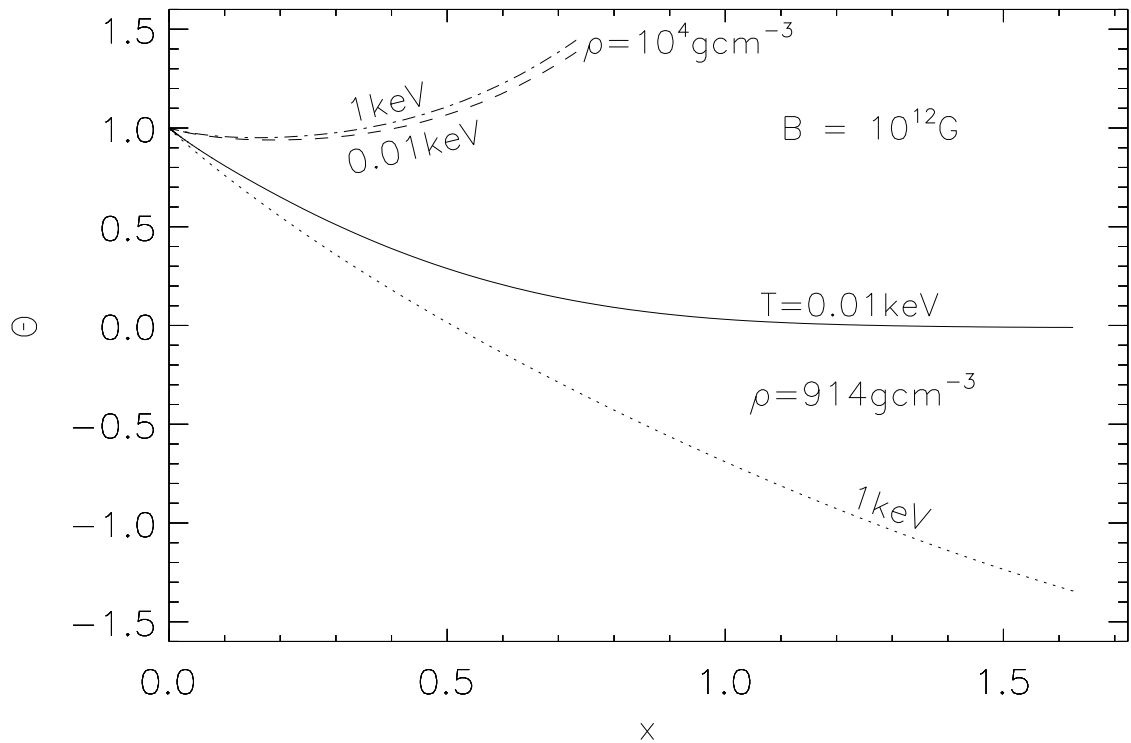}
\caption{
The potential $\Theta$ as a function of the radial coordinate $x$ for
two unit cells corresponding to two different values of the matter density.
Results are shown for two values of temperature and a magnetic field 
strength of $10^{12}$ G. The behaviour of $\Theta$ for $T = 0.01$ keV is very 
similar to the zero temperature case. Note that for higher temperatures, 
such as $T = 1$ keV shown here, $\Theta$ becomes negative at the low matter 
density whereas at the higher density it takes values which are higher than 
in the zero temperature case.
\label{Fig_Theta}}
\end{figure}

In practice, one solves (\ref{tfeq3}) with the initial condition 
(\ref{boundary4}) and a range of values for $\Theta'(0)$.  In each 
case one then determines $x_0$ from (\ref{boundary5}).  By varying 
$\Theta'(0)$ one obtains the necessary variation in the values of 
$x_0$ and hence of $\rho$ (see e.g.\ figure~\ref{Fig_Theta}).

The density and pressure profiles in an isothermal surface layer of 
a neutron star can be computed from the equation of state.
The surface layer, where TF theory can reasonably be 
assumed to apply in a typical neutron star, is so thin  
that the gravitational acceleration $g_{\rm s}$ can be taken as constant. 
Denoting by $z$ the depth from the surface, the equation of 
hydrostatic equilibrium for the pressure $P$ is
\beq
\frac{dP}{dz} = \frac{dP}{d\rho}\frac{d\rho(z)}{dz} = g_{\rm 
s}\rho(z)\label{hydro}.
\eeq
Inserting (\ref{ptf}) for $P$ one may obtain $\rho(z)$ by 
integration (see a more detailed discussion of isothermal atmospheres
in \S 4).  As in 
\cite*{FushikiEtAl89} and Paper I we find it more convenient, however,  to 
express the 
equilibrium condition in terms of the atomic chemical potential,
\beq
\mu_{\rm atom}=F_{\rm TF}+P_{\rm TF}v\label{muatom}
\eeq
where 
\beq
F_{\rm TF}=\int_{r\leq r_{0}}f_{\rm el}(n_{\rm TF}({\bf r})d^3{\bf r}
-\int_{r\leq r_{0}}\frac{n_{\rm TF}({\bf r})}{r}d^3{\bf r}+
\int_{r^{'}\leq r_{0}}\int_{r\leq r_{0}}\frac{n_{\rm TF}({\bf r})n_{\rm 
TF}({\bf r'})}{|{\bf r}-{\bf r}'|}d^3{\bf r}d^3{\bf r}'\label{Fatom}
\eeq
is the free energy per atom, and 
\beq
v=(4\pi/3)r_{0}^3
\eeq
is the volume per atom. The equilibrium condition is (\cite{FushikiEtAl89})
\beq
\mu_{\rm atom}(z)=\mu_{\rm atom}(0)+Am_{\rm n}g_{\rm s}z.
\label{muatom-z}
\eeq
If $\mu_{\rm atom}$ is known as a function 
of the density $\rho$, this immediately gives $\rho(z)$ and hence 
also $P(z)$ from the equation of state.

%
%
In our calculations
we have chosen to define the surface of the star to be where the matter
density, $\rho$, matches the lowest density, $\rho_0$, achieved in the
zero temperature case. For
each case of $T$ and $B$ used, we find the value of
$\mu_{\rm atom}(z=0) = \mu_{\rm atom}(\rho = \rho_0)$.
It is then easy to calculate the depth $z$ for each value of $\rho$
from equation (\ref{muatom-z}) and hence evaluate $\rho(z)$ and $P(z)$.
The results for isothermal atmospheres are shown in figures~9 and~10
and discussed further in \S 4.
It should be kept in mind that free atoms cannot exist at $P=0$ if $T>0$, so 
there is no sharp surface.

\subsection{Scaling Relations}

The equations (\ref{tfeq3}) and (\ref{boundary5}) have three parameters, 
$\beta=B/Z^{4/3}$,  $\tau=T/Z^{4/3}$
and $x_{0}=Z^{1/3}r_{0}$. By scaling they are equivalent to the original 
equations 
(\ref{tfeq2}) and (\ref{boundary3}) 
with four parameters, $Z$, $B$, $T$ and $r_{0}$. From (\ref{p}), 
(\ref{free}), (\ref{muatom}) and (\ref{Fatom}) we 
obtain the corresponding scaling relations for the pressure, the free energy 
and the atomic chemical potential. 
The matter density $\rho$ (cf.\ equation (\ref{rho})),
is a more convenient 
variable than $r_{0}$ and we write the scaling relations in the 
following way:
\beq
P_{\rm TF}(\rho,B,T;Z',A')=(Z'/Z)^{10/3}P_{\rm 
TF}((ZA/Z'A')\rho,(Z'/Z)^{4/3}B,(Z'/Z)^{4/3}T;Z,A)
\eeq
\beq
F_{\rm TF}(\rho,B,T;Z',A')=(Z'/Z)^{10/3}
F_{\rm TF}((ZA/Z'A')\rho,(Z'/Z)^{4/3}B,(Z'/Z)^{4/3}T;Z,A)
\eeq
\beq
\mu_{\rm atom}(\rho,B,T;Z',A')=(Z'/Z)^{7/3}
\mu_{\rm atom}((ZA/Z'A')\rho,(Z'/Z)^{4/3}B,(Z'/Z)^{4/3}T;Z,A)
\eeq
These relations allow us to compute the  pressure, free energy and 
chemical potential for any pair $(Z',A')$ if it is known for some 
reference pair $(Z,A)$ as e.g.\ for iron with $Z=26$, $A=56$.

\subsection{Limiting cases and validity}

There are three limiting cases, where TF theory 
simplifies and eventually passes over into the theory of a noninteracting 
electron gas. 
In terms of the parameters $\beta=B/Z^{4/3}$, $\tau=T/Z^{4/3}$ and 
$x_0=Z^{1/3}r_0$ these cases are:
\begin{itemize}
\item High density limit: $x_0(1+\beta)^{2/5}\ll 1$.
\item High temperature limit: $\tau/(1+\beta)^{2/5}\gg 1$.
\item Low density limit: $\tau\ln x_0/(1+\beta)^{2/5}\gg 1$.
\end{itemize}
To understand these limits one should recall that for an isolated 
TF atom at $T=0$ the radius $r_{B,Z}$ is of the order 
$Z^{-1/3}(1+\beta)^{-2/5}$  
and the ground state energy $E(B,Z)$ of the order 
$-Z^{7/3}(1+\beta)^{2/5}$ (\cite{FushikiEtAl92}). The high density limit 
corresponds 
therefore to $r_0\ll r_{B,Z}$, i.e.\ highly compressed matter. The high 
temperature limit means that $T\gg \vert E(B,Z)\vert/Z$, i.e.\ the thermal 
energy is much 
larger than the ground state energy per electron for isolated atoms. The 
low density limit is mainly of interest as a check for the correctness of 
the numerical computations.  In this limit the electron gas is essentially a 
classical ideal gas of density $n\sim r_0^{-1/3}$, because the chemical 
potential $\sim T\ln n$ of free electrons is much lower than 
the chemical potential $\sim E(B,Z)/Z$ of electrons bound to the nuclei.

In the three limiting cases 
TF theory may be compared with the simpler {\em uniform model} (see, 
e.g., \cite{FushikiEtAl89} and Paper I), 
defined by the free energy
\beq F_{\rm u}(\rho;B,T)={M\over \rho}f_{\rm
el}\left((Z/ M)\rho;B,T\right)-{9\over 10}Z^2\left({\rho\over 
M}\right)^{1/3},\label{free_uni}
\eeq
where $M=Am_{\rm n}$ and we use $\rho$ as a variable instead of
$r_0$. The mean electron density is $(Z/M)\rho$, and 
the first term in (\ref{free_uni}) is the free energy of a ball of radius 
$r_0$ of a uniform, 
noninteracting electron gas, whereas the second term is the 
Coulomb energy of the ball, due to electronic repulsion and attraction of 
the nucleus. The pressure in the uniform model is 
\beq
P_{\rm u}(\rho;B,T)=P_{\rm el}\left((Z/ M)\rho;B,T\right)-{3\over 
10}\left({4\pi\over 3}\right)^{1/3}
Z^2 \left({\rho\over M}\right)^{4/3},\label{press_uni}\eeq
with $P_{\rm el}$ given by (\ref{p}).
Since $P_{\rm el}$ behaves like $\rho^{5/3}$ at high $\rho$ and as 
$\rho T$ at small $\rho$ or high $T$ it is also clear that the first 
term in (\ref{press_uni})
dominates over the second in the limiting cases considered. 
A comparison of $P_{\rm TF}$ with $P_{\rm u}$ and $P_{\rm el}$ is shown in 
figure~6.

In \cite*{BH96}
the status of TF theory with a magnetic field at nonzero 
temperature as an approximation of quantum mechanics is investigated, 
generalizing the zero temperature limit theorems of \cite*{LSY94}.
It is found that  
thermodynamical functions calculated from quantum mechanics converge to the 
corresponding functions in TF theory in the limit when the nuclear charge 
$Z$ and the magnetic field $B$ 
tend to infinity, provided $B/Z^3\ll 1$. The temperature $T$ and  
the radius $r_0$ are in this limit 
scaled in such a way that $\tau/(1+\beta)^{2/5}$ and $x_0(1+\beta)^{2/5}$
are fixed, i.e.\ $T$ scales as $Z^{4/3}(1+(B/Z^{4/3}))^{2/5}$ and $r_0$ as
$Z^{-1/3}(1+(B/Z^{4/3}))^{-2/5}$. In compressed matter and/or at high 
temperature the condition $B/Z^3\ll 1$ can be relaxed. In fact by 
the heuristic argument in Paper I, TF theory for compressed matter at
zero temperature, 
i.e.\ for $r_0<Z^{-1/3}(1+(B/Z^{4/3}))^{-2/5}$,  can be expected to be a 
good approximation, if $Z\gg 1$ and $B\ll Z/r_0^2$. If $T\gg B$, many Landau 
levels are excited, and the magnetic 
field becomes irrelevant. Hence at such extreme temperatures the condition 
for validity is simply $Z\gg 1$. Summing up, the domain of validity of TF 
theory at non-zero temperature should be at least as large as that for zero 
temperature illustrated in Fig.~1 in Paper I, and strictly larger for $T> B$.


%
%
\section{Numerical Methods}
\label{sec_numcalc}


Since the calculations presented in this article are an extension of
those presented in Paper I, the numerical methods are in most
aspects similar. The details given in that paper will not be repeated,
but here we present
a general outline together with considerations specific to the
case of non-zero temperatures.

To solve the TF equation (\ref{tfeq2}) numerically, we use a
fourth-order Runge-Kutta method with adaptive stepsize control
(\cite{PressEtAl92}) with the boundary conditions (\ref{boundary4}) and
(\ref{boundary5}).  First we pick a suitable spectrum of values of
$\Theta'(0)$ and then integrate outwards from $\Theta(0)$ until
condition (\ref{boundary5}) is satisfied.  We then know both $x_{0}$ and
$\Theta(x_{0})$ which we can use to calculate the density and the chemical
potential, and hence the
corresponding pressure for a given temperature and magnetic field.

It is worth mentioning some changes from the zero temperature case.  The kinetic
energy density, denoted by $w$ in Paper I, is replaced with the free energy density,
$f$. Whereas $w$ is always non-negative, the free energy density can
become negative because of the added temperature term.  This allows
the electrochemical potential $\Theta$ to become negative for high
temperatures and low densities, since in such cases the free energy,
$f$, can have a negative derivative with respect to $n$, yielding $\mu
< 0$. An example of this is seen in figure~\ref{Fig_Theta} where we
plot $\Theta$ as a function of $x$ for two different densities and
temperatures with $B$
fixed.  Also, when $T>0$ there are no free atoms at $P=0$,
contrary to the case for zero temperature (e.g.\ Paper I).


As in Paper I, we chose to include the first 100 terms of the infinite
sums in the formulas (\ref{pdash}) and (\ref{p}) for $n$ and $P$
ensuring a relatively high precision without too much computational
cost.
Here, we encountered a similar divergence problem as in the zero
temperature case; close to the
nucleus, both $n$ and $\mu$ tend to infinity.
We also found the
convergence of the sums in (\ref{pdash}) and (\ref{p}) to be slow for
magnetic fields low compared to the temperature.
Both problems may be solved by
using the approximations (\ref{limit1}) and (\ref{limit2}), since in
the first case $\xi \ll |\eta|$ and in the second $\xi \ll 1$. It then
remains to find a criteria for where to switch over to these
approximations. It is convenient to rewrite the expressions depending on $\mu$
in equations (\ref{pdash}) and (\ref{p}) as
$$
\left[ I_{k} \left(\xi \zeta
\right) + 2 \sum_{\nu =1}^\infty I_{k} \left(\xi (\zeta - \nu )
\right)\right]
$$
with
$$
\zeta:=\frac { \mu}{B}
$$
and determine in terms of $\zeta$ and $\xi$ where to switch over to the
approximations to obtain the desired accuracy.
We decided that in the worst case the relative error in
$P$ and $n$ should never exceed $2\times 10^{-4}$ for any value of
$\mu$ at a given temperature and magnetic field strength.
We calculated this error for the two methods, using only 100 Landau
levels on 
one hand and the approximations on the other and comparing the results
with the values obtained with 5000 Landau levels (which we consider as
exact). This comparison yielded
critical values for $\zeta$, referred to as $\zeta_{\rm c}$, above which the
approximations (\ref{limit1}) and (\ref{limit2}) are used instead
of the partial sum over 100 levels (see table \ref{Table_zeta}). Since
different
Fermi-Dirac integrals enter the calculations of $n$ and $P$, slightly
different $\zeta_{\rm c}$ values are appropriate in each case for the same
$\xi$.
The Fermi-Dirac integrals were evaluated using the method of 
\cite*{CodyThatcher67}
which assures a relative error less than $3 \times 10^{-9}$. (Note that
in their paper, the value of $q_{s}$ for $s=0$ and $n=4$ in table IIC 
should read 1.00 and not 1.05.)


%
%
\section{Results}
\label{sec_results}
In this section we present the results of our numerical TF calculations for
bulk matter in high magnetic fields and at finite temperature. We remind the 
reader that our reference element is iron with $Z = 26$ and  $A = 56$ and
that the treatment is non-relativistic. Hence the highest density used in our 
calculations is $10^{6}~{\rm g~cm}^{-3}$ and the highest temperature is 
$100$ keV ($1.16 \times 10^{9}$ K). As in Paper I the range in magnetic field 
strength is from $10^{10}$ to $10^{13}$ G. 
The reason for this is that in the case 
of iron one obtains the zero-field results (see e.g.\ \cite{AbrahamsShapiro91}) if the fields are much weaker than $10^{10}$ G 
whereas for field strengths slightly higher than $10^{12}$ G essentially 
only the lowest Landau level is populated and one obtains the results of
\cite*{ConstantinescuMoruzzi78} and \cite*{AbrahamsShapiro91}.  

Our presentation of results is mainly in graphical form but tabulated results 
are available at the website http://www.raunvis.hi.is/\~{}ath/TFBT. By use of the scaling relations 
discussed in \S 2.5 the numerical TF results can also be used for elements 
other than iron and for different temperatures and magnetic field strengths. 

\begin{figure}
\plotone{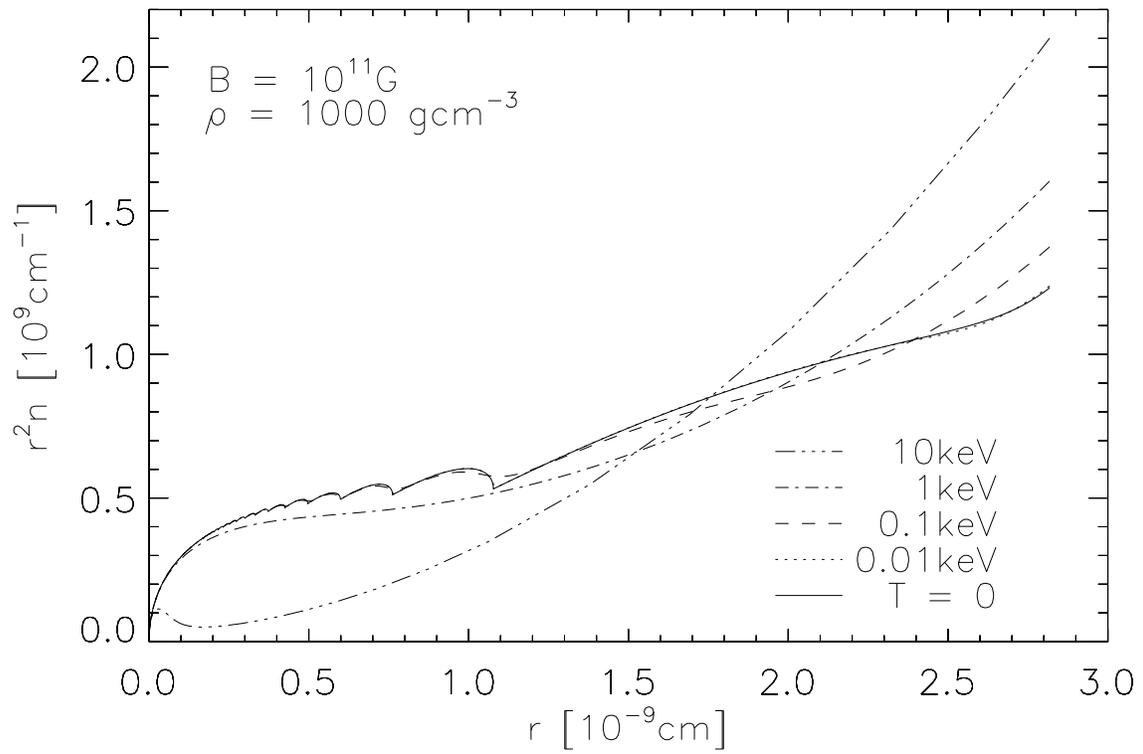}
\caption{
The quantity $r^{2} n$ for a spherical unit cell of iron in the
TF approximation. The matter density is $1000~{\rm g~cm}^{-3}$ and the 
magnetic field strength $10^{11}$ G. Results are shown for five different 
values of the temperature.
\label{Fig_r2n}}
\end{figure}

Figure~\ref{Fig_r2n} shows a typical example of how temperature changes 
the electron distribution inside a unit iron cell of a given size, 
corresponding to fixed matter density. The quantity plotted is $r^{2} n(r)$ 
which is useful since the the number of electrons in a shell is proportional 
to the area under the corresponding curve. The case shown is 
for a field strength 
of $10^{11}$ G and a density of $1000~\rm{g~cm}^{-3}$. The main thing 
to notice is that thermal effects tend to smooth out the distinct Landau shell 
structure which is so prominent in the zero temperature case (see
\cite{FushikiEtAl92}, and Paper I). However, as long as the temperature
is lower than
approximately one tenth of the cyclotron energy, $\hbar \omega_{B}$
($\approx 11.58 B_{12}$ keV where $B_{12} = B/10^{12}$ G), thermal 
effects are not particularly pronounced. It is only when the temperature 
is comparable to or higher than the cyclotron energy ($\approx 1$ keV for 
$10^{11}$ G) that the electron distribution becomes 
drastically different from the zero temperature case. This is a reflection
of the temperature dependence of the relation between $n$ and $\mu$ for the 
free electron gas shown in figure~\ref{Fig_nmu}. From
figure~\ref{Fig_r2n} it is also clear that 
the number density of electrons at the edge of the unit cell increases 
with increasing temperature and thus the TF pressure of bulk matter at given 
matter density and field strength also increases with temperature.

In figures~4 to~10 we present the results of our calculations of the bulk 
properties of dense matter. Figure~4 shows the TF equation of state for iron, 
$P_{\rm TF}$ versus $\rho$, where $P_{\rm TF}$ is 
given by equation (\ref{tfeq4}) and $\rho$ by equation (\ref{rho}). 
Each curve corresponds to a
fixed value of the temperature. The presence of a Landau shell structure in 
the electron distribution is reflected in a clear oscillatory nature of the 
pressure-density relationship as long as the temperature is lower than
approximately one tenth of the cyclotron energy (i.e.\
$T \leq 1 B_{12} {\rm keV}$). For somewhat higher temperatures the oscillations 
are smoothed out as previously noted by \cite*{GadiyakEtAl81}.

In figure~\ref{Fig_eosB12} we compare the TF equation of state for a field strength of
$10^{12}$ G with that of a nondegenerate gas of free electrons 
with uniform density $n = Z \rho / A m_{\rm n}$ and temperature $T$ but in 
zero magnetic field. This pressure is given
by equation (\ref{idealgas}) or equivalently by
\begin{equation} 
P_{\rm class} = P_{\rm class}(\rho, T, B = 0) = Z \rho k T / A m_{\rm
n}~ ,
\label{resPclass}
\end{equation}
where we use the notation $P_{\rm class}$ instead of $P_{\rm el}$ in order to avoid
confusion with the more general expressions for $P_{\rm el}$ in \S 2.3 (also note
the change of units from those used in \S 2). We have
chosen the value of $10^{12}$ G for this particular discussion but the
same qualitative features are present for all $B$.
Figure~\ref{Fig_eosB12} shows that 
at low densities the magnetic TF pressure  approaches $P_{\rm class}$
for fixed $B$ and $T$. 
The same is true in the limit of high temperatures at fixed $\rho$ and $B$.
Hence at very low densities and/or very high temperatures the TF pressure 
becomes independent of $B$ as expected from the discussion of asymptotic 
behaviour in \S 2.6. 
On the other hand at high enough densities and as long as the temperature
does not become too high both thermal and magnetic effects 
on the equation of state become negligible and it approaches the non-magnetic 
TF equation of state of iron at zero temperature (Paper I). This limiting 
behaviour is clearly seen in figure~\ref{Fig_eosT} which shows $P_{\rm
TF}$ versus $\rho$ for $T = 0.1$ keV and several values of the
magnetic field.

In figure~6 we compare the TF pressure, $P_{\rm TF}$, with $P_{\rm u}$, the pressure
of the uniform model discussed in \S 2.6 (see equation (\ref{press_uni})). Results are
shown for a magnetic field with strength $10^{12}$ G and two values of
the temperature. Also shown is the pressure of a noninteracting
electron gas, $P_{\rm el}$, given by equation (\ref{p}). It is clear from
the figures that the three models have the same behaviour in the limit
of high or low densities and at high temperatures as already discussed
in \S 2.6. The differences between the different approximations are
greatest at low temperatures and in a restricted range of relatively low
densities. This is approximately the range in which one expects
contributions from the thermal motion of the nuclei to be important
(see the discussion below).

For completeness we show in figure~7 the Thomas-Fermi free energy for iron in 
bulk, $F_{\rm TF}$, as a function of $\rho$ for selected values of
$T$ and $B$ (see equation (\ref{Fatom})).

At this point we would like to mention a few corrections to the 
equation of state that go beyond the TF approximation and are therefore 
not included in our calculations. For further discussion of this and
related topics we refer the reader to Paper I and the papers by
\cite*{FushikiEtAl92} and \cite*{YakovlevKaminker94}.

As already mentioned in the introduction the pure TF method neglects both
exchange effects between the electrons as well as ``gradient corrections'' of the
von Weizs{\"{a}}cker type. Although these effects are not expected to make much 
difference as far as bulk properties of matter are concerned 
(\cite{FushikiEtAl92}) it should be kept in mind that they may become important
at very low temperatures and the densities corresponding to the surface 
densities of neutron stars (\cite{FushikiEtAl89}; \cite{AbrahamsShapiro91}; 
Paper I). To investigate this in detail requires more exact 
calculations than are provided by the one dimensional (spherical) TF method.

An additional simplification is that in TF theory the nuclei are 
assumed to be stationary at fixed relative distances determined 
by the matter density, 
their number density given by $n_{\rm i} = \rho / A m_{\rm n}$. 
Thus the thermal motion of the nuclei is neglected in calculations of the 
TF pressure. This motion is only weakly affected by the
magnetic field and for most applications of relevance to the investigation
of neutron star envelopes it is a
fairly good approximation to regard the nuclei as forming an
ideal non-degenerate gas with pressure
\begin{equation}
P_{\rm i} = n_{\rm i}kT = \rho k T / A m_{\rm n}
\label{resPi}
\end{equation}  
(for a more detailed  discussion see e.g.\ \cite{GudmundssonEtAl83}, and 
\cite{YakovlevKaminker94}). Note however that $P_{\rm i} = P_{\rm class}/Z$, 
where $P_{\rm class}$ is the 
pressure of the ideal non-degenerate gas of electrons at the same
$\rho$ and $T$ (see equation (\ref{resPclass})) 
and thus the pressure of the nuclei is small compared to the TF
pressure except at low temperatures and and low densities, which are
of little importance for compressed matter.
This can e.g.\ be seen by inspecting
figures~\ref{Fig_eosB12} and~6 which show a 
comparison between $P_{\rm TF}$, $P_{\rm class}$, $P_{\rm el}$,
and $P_{\rm u}$ for $B = 10^{12}$ G.
We thus feel justified in neglecting the thermal
motion of the nuclei in our calculations. 
We point out however that in case one wants
to include the contribution of the nuclei to the total pressure and chemical
potential this can easily be done by adding them to the TF results, thus 
obtaining
\begin{equation}
P = P_{\rm TF} + P_{\rm i}
\label{resP}
\end{equation}
and
\begin{equation}
\mu = \mu_{\rm atom} + \mu_{\rm i}~ ,
\label{resMu}
\end{equation}
where $\mu_{\rm atom}$ is given by equation (\ref{muatom}). For an ideal non-degenerate
gas of nuclei in zero magnetic field $P_{\rm i}$ is given by equation 
(\ref{resPi}) above and
\begin{equation} 
\mu_{\rm i} = kT~ {\ln{[n_{\rm i} (2 \pi \hbar^{2}/m_{\rm n}kT)^{3/2}]}}.
\end{equation}
A more detailed treatment taking into account the non-ideal behaviour of 
nuclei due to Coulomb interactions between nuclei and quantum effects 
is beyond the scope of this paper. We would however like to remind the reader
that these effects are expected to make only relatively small
corrections to the equation of state of bulk matter for most of the 
temperatures and densities of interest in the study of neutron star envelopes. 
On the other hand they do have a considerable influence on the transport 
properties in magnetic neutron stars, and they also determine which form
or phases of matter are present in the envelope (see e.g.\ \cite{GudmundssonEtAl83};
\cite{VanRiper88}; \cite{YakovlevKaminker94}; \cite{LaiSalpeter97},
and references therein).

We now return to the discussion of the results of our TF
calculations. Figure~\ref{Fig_cs} shows the isothermal sound velocity,
\begin{equation}
a_{T} = \left({\frac{\partial P}{\partial \rho}}\right)^{1/2}_T
\label{resCs}
\end{equation}
with $P = P_{\rm TF}$, as a function of density for several values of the
temperature and a magnetic field of $10^{12}$ G.
This quantity is related to the adiabatic sound velocity $a_{\rm s} = (\partial
P/\partial \rho)_s^{1/2}$ by
$$
  a_T^2 = \frac{c_v}{c_p} a_{\rm s}^2,
$$
where $c_p$ and $c_v$ are the specific heats at constant pressure and
volume, respectively.
As discussed in
Paper I (see also \cite{LandauLifshitz60}) $a_{\rm s}$ is the velocity of
longitudinal magnetosonic waves when the wavevector is parallel to the
magnetic field. Figure~\ref{Fig_cs} shows that the oscillatory behaviour which is
due to the Landau
shell structure in the electronic density distribution of the unit
cells is clearly present in $a_T$ at finite temperatures. Even though
thermal effects tend to smooth them out they are quite pronounced as
long as the temperature is less than approximately one tenth of the
cyclotron energy. It is only when the temperature is of order $10~
B_{12}~ {\rm keV}$ or higher that the oscillations disappear
completely, an effect that has already been discussed in connection
with the equation of state.  In two or three dimensional TF
calculations an additional smoothing effect would probably be present
due to the non-spherical nature of the unit cells. However as argued in
Paper I they would not be big enough to wipe out the oscillations.

These de Haas-van Alphen type oscillations are reflected in the
isothermal compressibility
of bulk matter $\kappa_{\rm bulk}$, which is related to the sound velocity by
\begin{equation} 
\kappa_{\rm bulk} = \frac{1}{\rho a_{T}^{2}}~ ,
\end{equation}
and hence they are also present in the density 
and pressure profiles of matter in the surface 
layers of neutron stars (Paper I). In most cases of interest the temperature 
in the surface layers rises quite steeply as a function of depth and hence the 
oscillatory behaviour in the matter density can not be investigated in detail 
without solving the thermal structure equations of the envelope.
This requires knowledge of the opacity as well as
the equation of state (\cite{GudmundssonEtAl83}; \cite{VanRiper88}). 
Some insight may clearly be gained however by considering
the simplified case of isothermal surface layers, an approximation which may 
in fact be appropriate for sufficiently old and relatively cold neutron stars.
Figures~9 and~10 show the results of such isothermal structure calculations
for a range of temperatures and two values of the magnetic field strength.
The method used in the calculations is that described at the end of \S 2.4.
The oscillatory behaviour is clearly seen in figure~9 as steps in the density
profiles for the lowest temperatures whereas the effect is somewhat weaker 
in the pressure profiles in figure~10. Notice also that at high temperature
both the density and pressure gradients are very small 
in the outermost layers. This is a result of the fact that 
there is no sharp boundary at $T>0$.
For this reason the
depth shown in figures~9 and~10 is measured from the point where the surface 
would be at zero temperature (see the discussion at the end of \S 2.4
and Paper I).  

In some applications it is also of interest to know the
column density, $\sigma$, either as a function of $z g_{\rm s}$ or of $\rho$.
In either case it is easy to calculate the 
column density from the surface to the
point in the star where the pressure is $P$ since it is simply given by the 
expression
\begin{equation}  
\sigma - \sigma_{\rm s} = \int_{0}^{z} \rho dz^{\prime}
                        = \int_{P_{\rm s}}^{P}
                        \frac{dP^{\prime}}{g_{\rm s}}
                        = (P - P_{\rm s})/g_{\rm s}~ ,
\end{equation}
where we have used equation (\ref{hydro}) and
$\sigma_{\rm s}$ and $P_{\rm s}$ are the column density and
pressure at the surface of the neutron star, respectively.
Hence the results shown in figures~4 and~10 can easily be used for evaluating
$\sigma$. 

We end the discussion of our results by estimating the effects of the thermal 
motion of the nuclei on the velocity of sound and hence also on
$\kappa_{\rm bulk}$ and the isothermal
surface structure. If we assume as before that the nuclei form 
an ideal non-degenerate gas at
temperature $T$ then by use of equations (\ref{resPi}) and 
(\ref{resCs}) we find that the TF sound velocity shown in figure~\ref{Fig_cs}
and denoted by $a_{T}$ has to be replaced by $a_{T}^{'}$ given by
\begin{equation}
a_{T}^{'} = (a_{T}^{2} + kT/Am_{\rm n})^{1/2}~ .
\end{equation}
By comparing the nuclear contribution, 
$(kT/Am_{\rm n})^{1/2} \approx 4 \times 10^{6}~ (T/{\rm keV})^{1/2}~ {\rm cm/s}$, to
the values of $a_{T}$ shown in figure~\ref{Fig_cs} one can see as before that it
is only at the lowest temperatures and in a restricted range of low densities 
that the thermal motion
of the nuclei needs to be taken into account. In particular it is clear that 
this motion has a negligible effect on the high density oscillatory 
behaviour which occurs at low temperatures.

\section{Conclusions}
\label{sec_conclusions}

In this paper we have presented the results of a detailed calculation of
the properties of hot dense matter in high magnetic fields using the 
temperature dependent Thomas-Fermi method and taking all Landau levels
into account.
By including temperature these calculations extend our previous work on the 
properties of cold dense matter in the TF approximation presented in Paper I.
 
On the grounds discussed in \cite*{FushikiEtAl92} and Paper I we expect 
that TF theory catches the main features of bulk matter composed of heavy 
elements like iron. This applies also for finite temperatures and high 
magnetic fields up to at least a few times $10^{12}$ G.

We have investigated the range of validity of the finite temperature TF model
for matter in a magnetic field and estimated the effects of the thermal 
motion of nuclei on the properties of matter in bulk. We find that the 
nuclear contribution to the total pressure and other bulk properties 
is very small except in a restricted range of low densities and
low temperatures.

Within the range of validity of the finite temperature TF theory we have
found that the Landau shell structure which is so pronounced in iron at zero 
temperature is also present at finite temperature as long as it
does not exceed about one tenth of the cyclotron energy. This structure is 
reflected in an oscillating component of the de Haas-van Alphen type in all 
the bulk properties of matter as functions of depth in isothermal surface 
layers of magnetic neutron stars. At temperatures around the cyclotron energy or
higher, thermal effects smooth out the oscillations. Further investigation of 
this problem including the additional effects of smoothing of oscillations 
due to the non-spherical shape of the unit cells requires 
two or three dimensional TF calculations. 
Furthermore an investigation of the thermal structure of
the surface layers when the assumption of isothermality is not valid requires
solving the thermal structure equations of neutron star envelopes. This
in turn requires knowledge of the transport properties of matter in high 
magnetic fields in addition to the equation of state presented in this work. 
Such study is beyond the scope of this paper. Finally we remind
the reader that our numerical results are available in tabular form
at the website http://www.raunvis.hi.is/\~{}ath/TFBT.

%

%

We are grateful to Chris Pethick and Sasha Potekhin for valuable discussions.
A.~Th.\  and E.~H.~G.\ would like to thank Nordita for generous
hospitality. J.~Y.\ acknowledges a grant from the Adalsteinn Kristj\'{a}nsson 
Foundation of the University of Iceland.
This work was partially supported by the Research Fund of the University of 
Iceland and by Danmarks Grundforskningsfond through its
support of the Theoretical Astrophysics Center.
%
%
%
\newpage

\bibliographystyle{haabib}
\bibliography{tfeos}

\newpage


\begin{figure}
\figurenum{4a}
\plotone{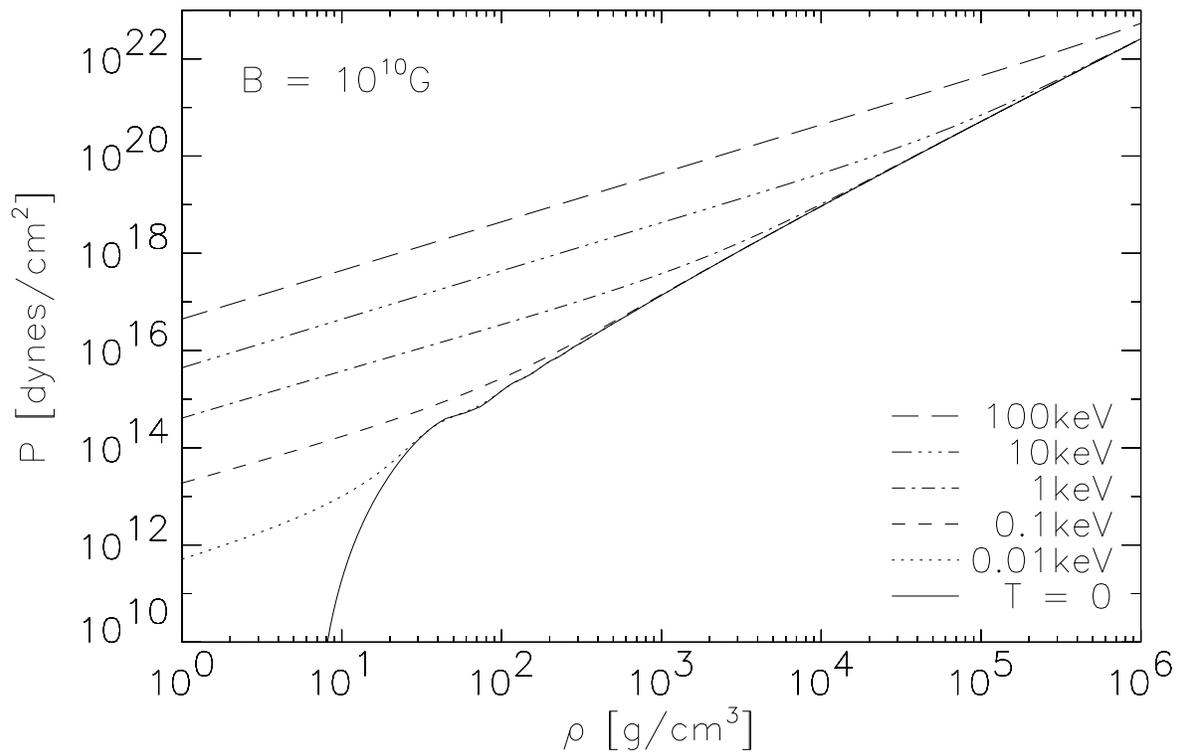}
\caption{
The Thomas-Fermi equation of state for iron in a magnetic field.
In each case results are shown for six values of the temperature, including
the zero temperature case. (a) $B = 10^{10}$ G. (b) $B = 10^{11}$ G.
(c) $B = 10^{12}$ G. (d) $B = 10^{13}$ G.  In 4c the pressure
of an ideal non-degenerate gas of free electrons in zero field, 
$P_{\rm class}$, is superimposed on the TF results for comparison. 
See the text for further explanation.
\label{Fig_eos}}
\end{figure}

\begin{figure}
\figurenum{4b}
\plotone{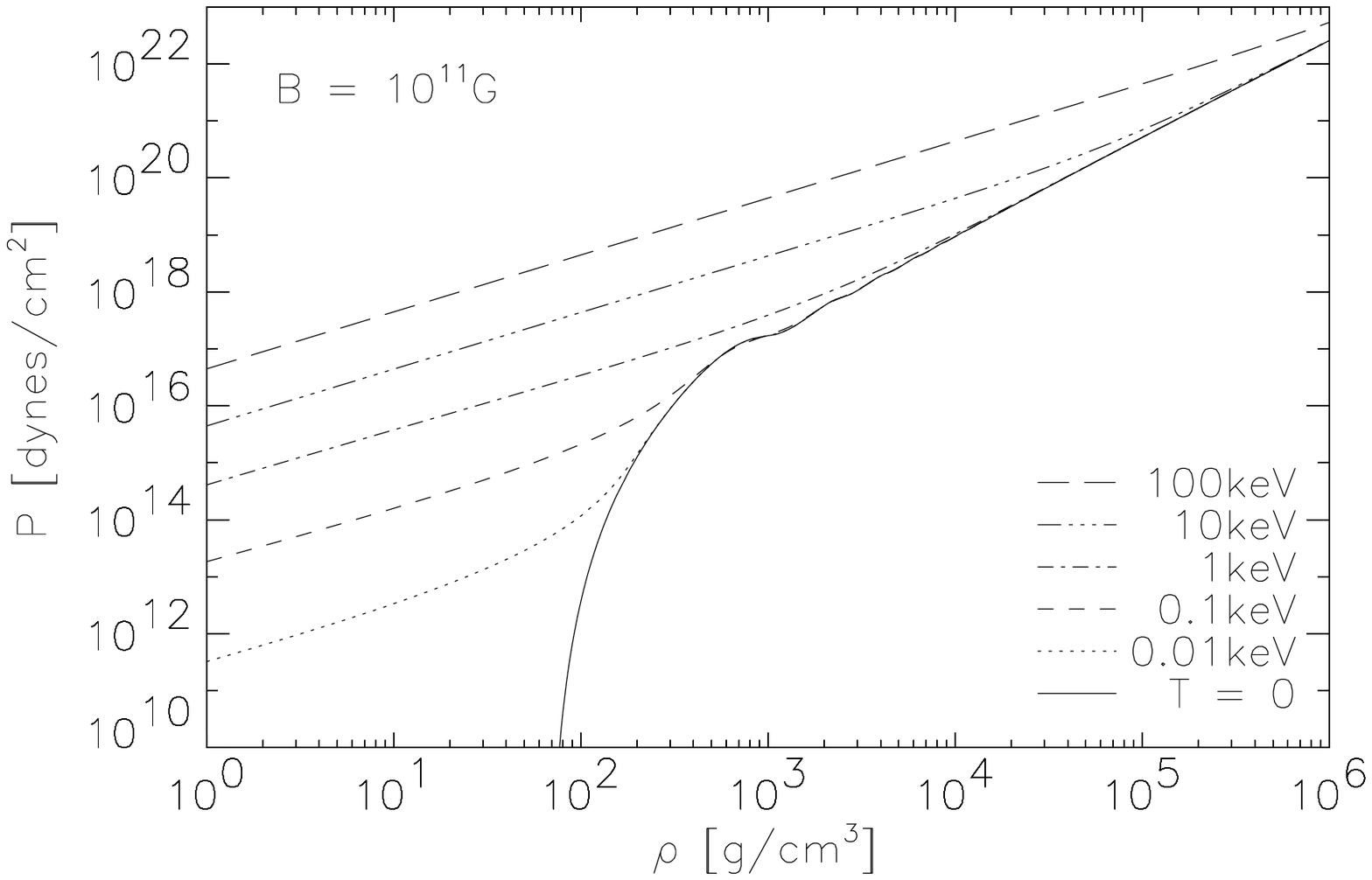}
\caption{}
\label{Fig_eosB11}
\end{figure}

\begin{figure}
\figurenum{4c}
\plotone{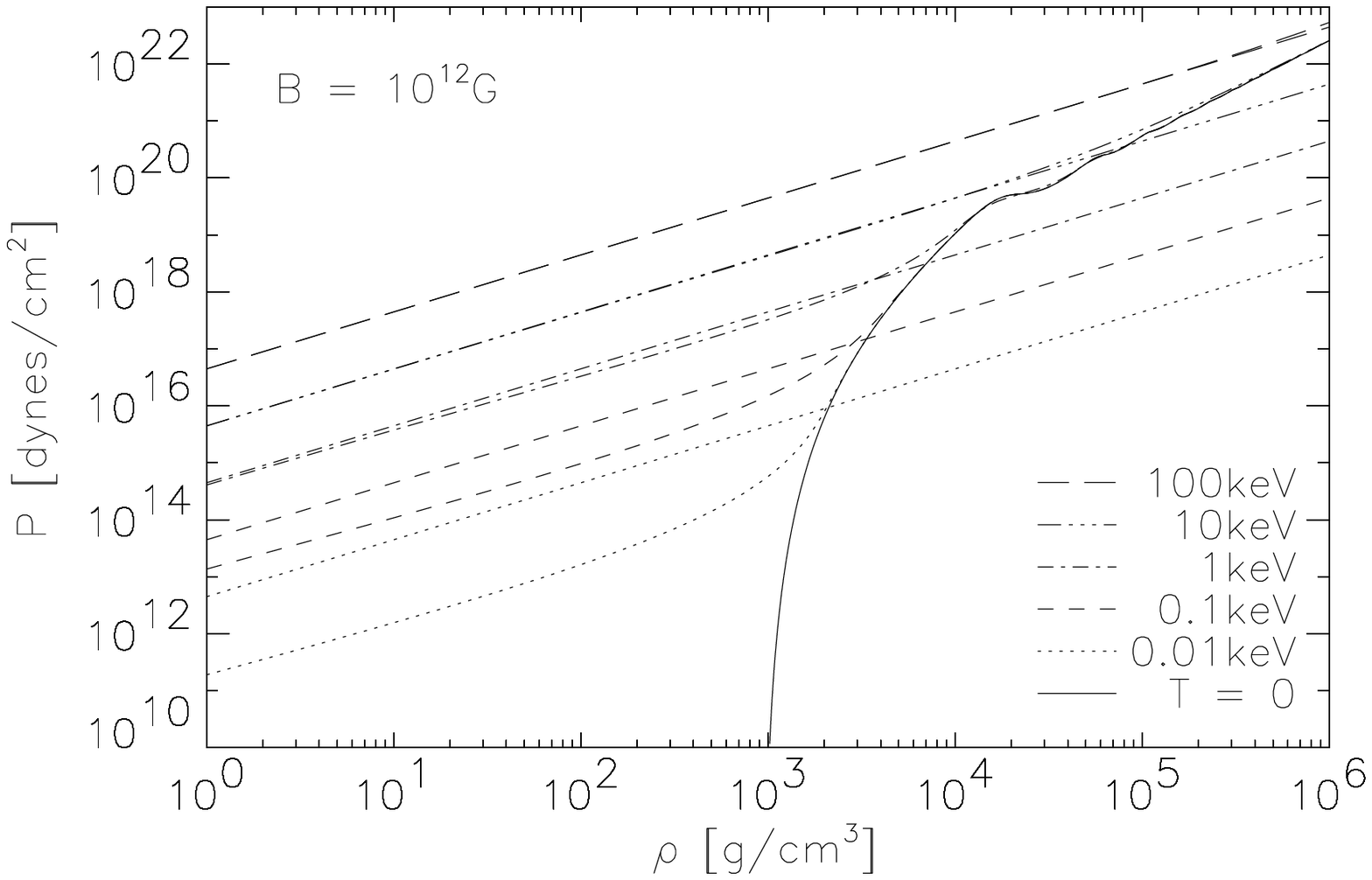}
\caption{}
\label{Fig_eosB12}
\end{figure}

\begin{figure}
\figurenum{4d}
\plotone{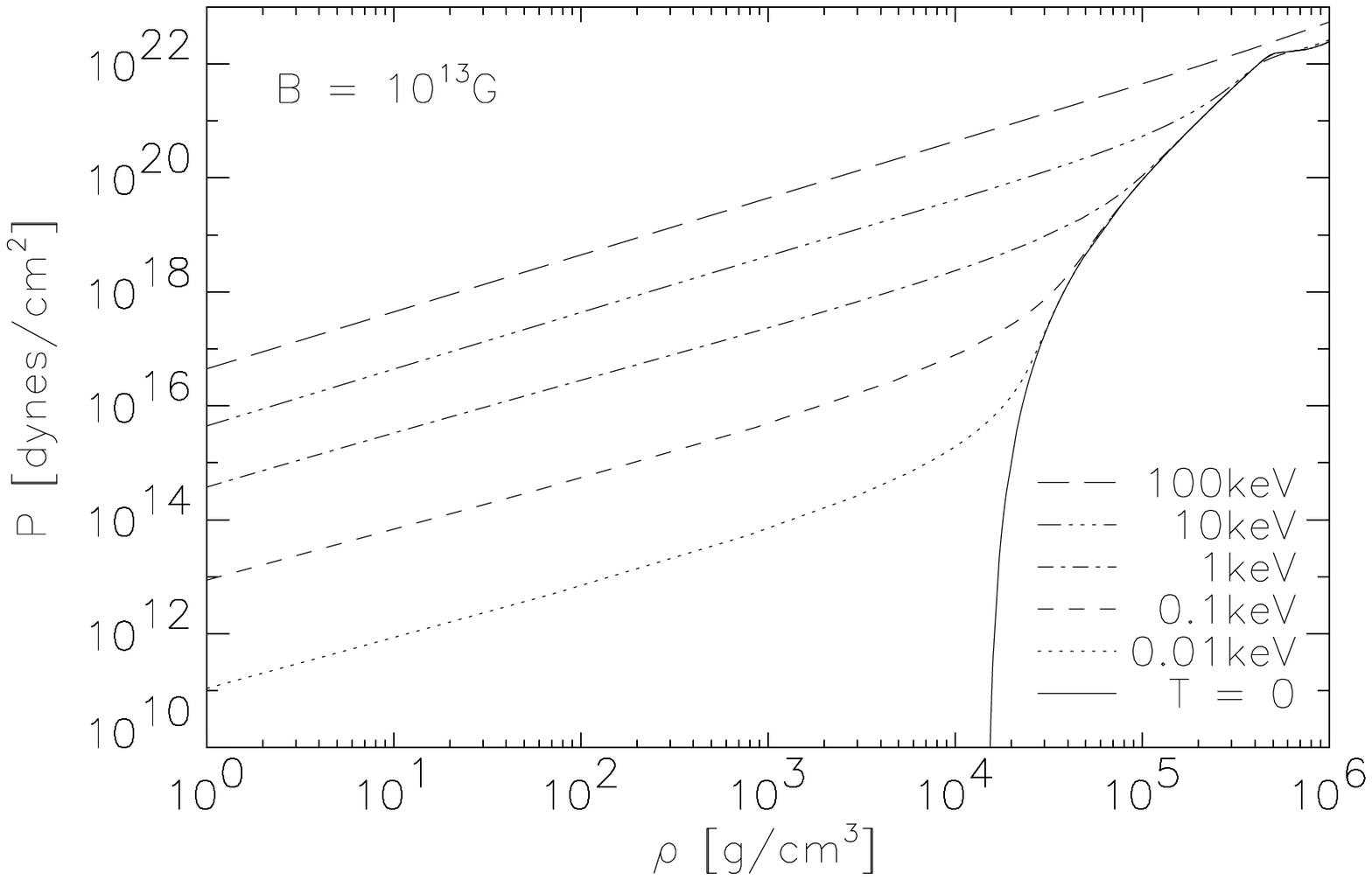}
\caption{}
\label{Fig_eosB13}
\end{figure}

\begin{figure}
\figurenum{5}
\plotone{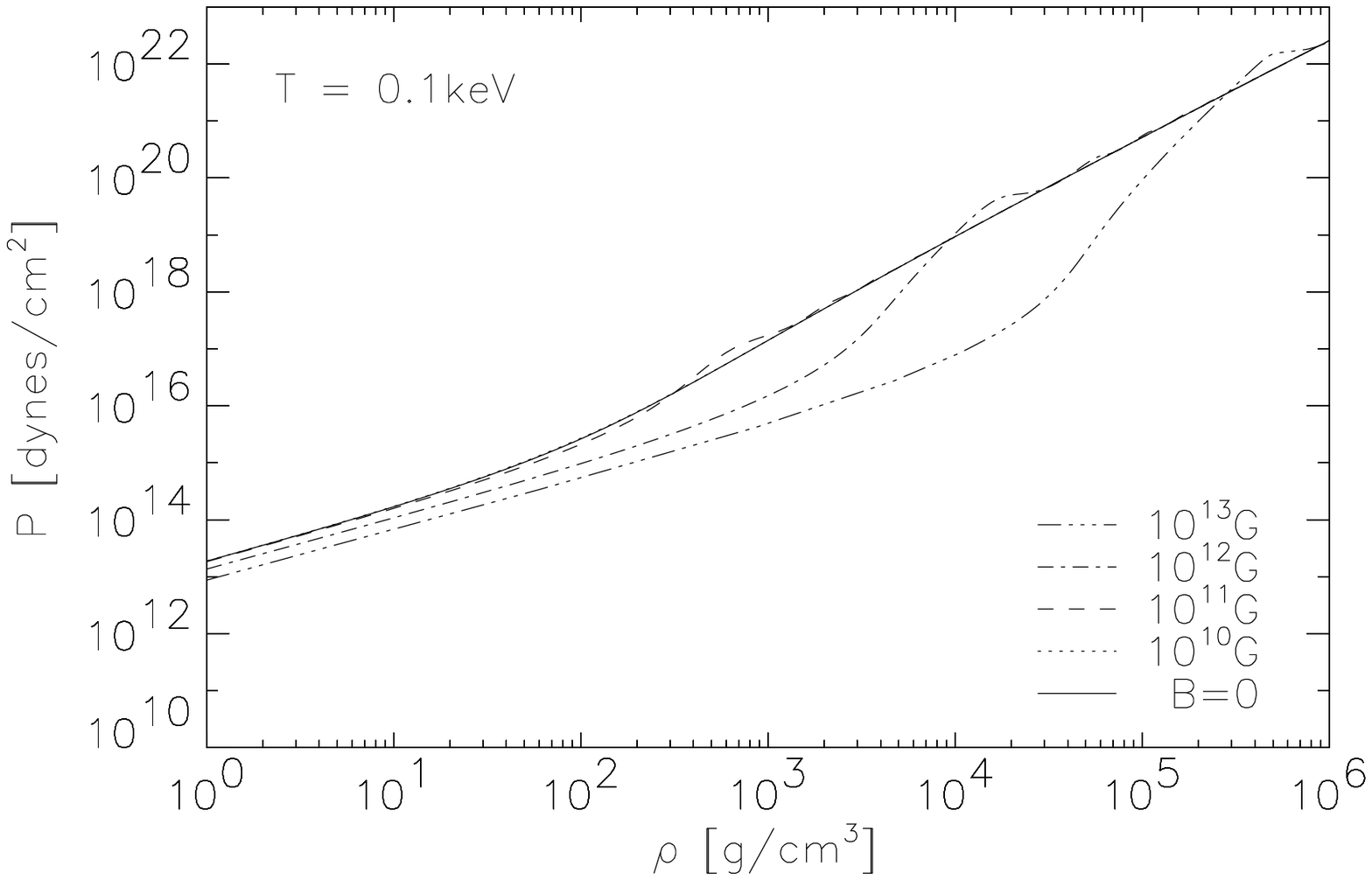}
\caption{
The Thomas-Fermi equation of state for iron at a temperature of 
$0.1$ keV and several values of the magnetic field strength. 
Note in particular that the curve for
$B = 10^{10}$ G falls almost completely on top of the zero field curve.
\label{Fig_eosT}}
\end{figure}

\begin{figure}
\figurenum{6a}
\plotone{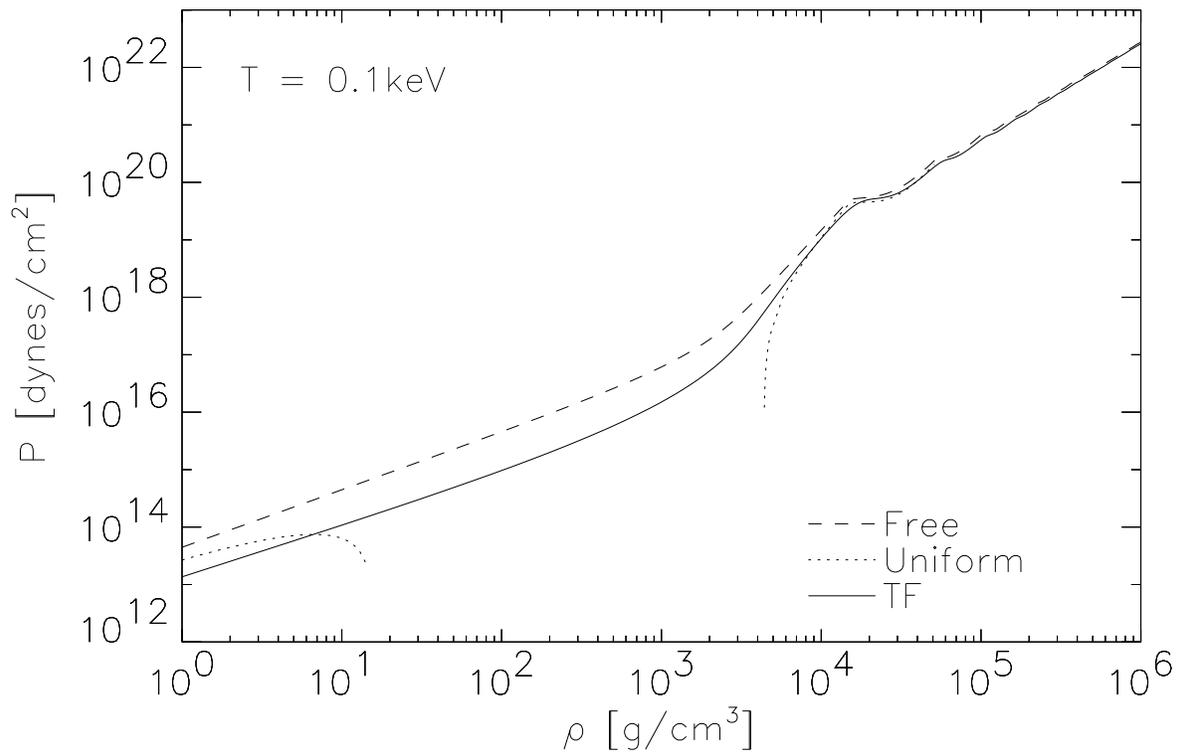}
\caption{
The equation of state for iron in three different approximations.
The TF pressure, $P_{\rm TF}$, is shown by the solid line. The
pressure of a Fermi gas of free electrons, $P_{\rm el}$, is shown by the dashed
line and the pressure in the uniform model, $P_{\rm u}$, by the dotted line. The
strength of the magnetic field is $10^{12}$ G. (a) $T = 0.1$ keV. (b) $T = 1$ keV.}
\label{Fig_PTF-uni-el}
\end{figure}

\begin{figure}
\figurenum{6b}
\plotone{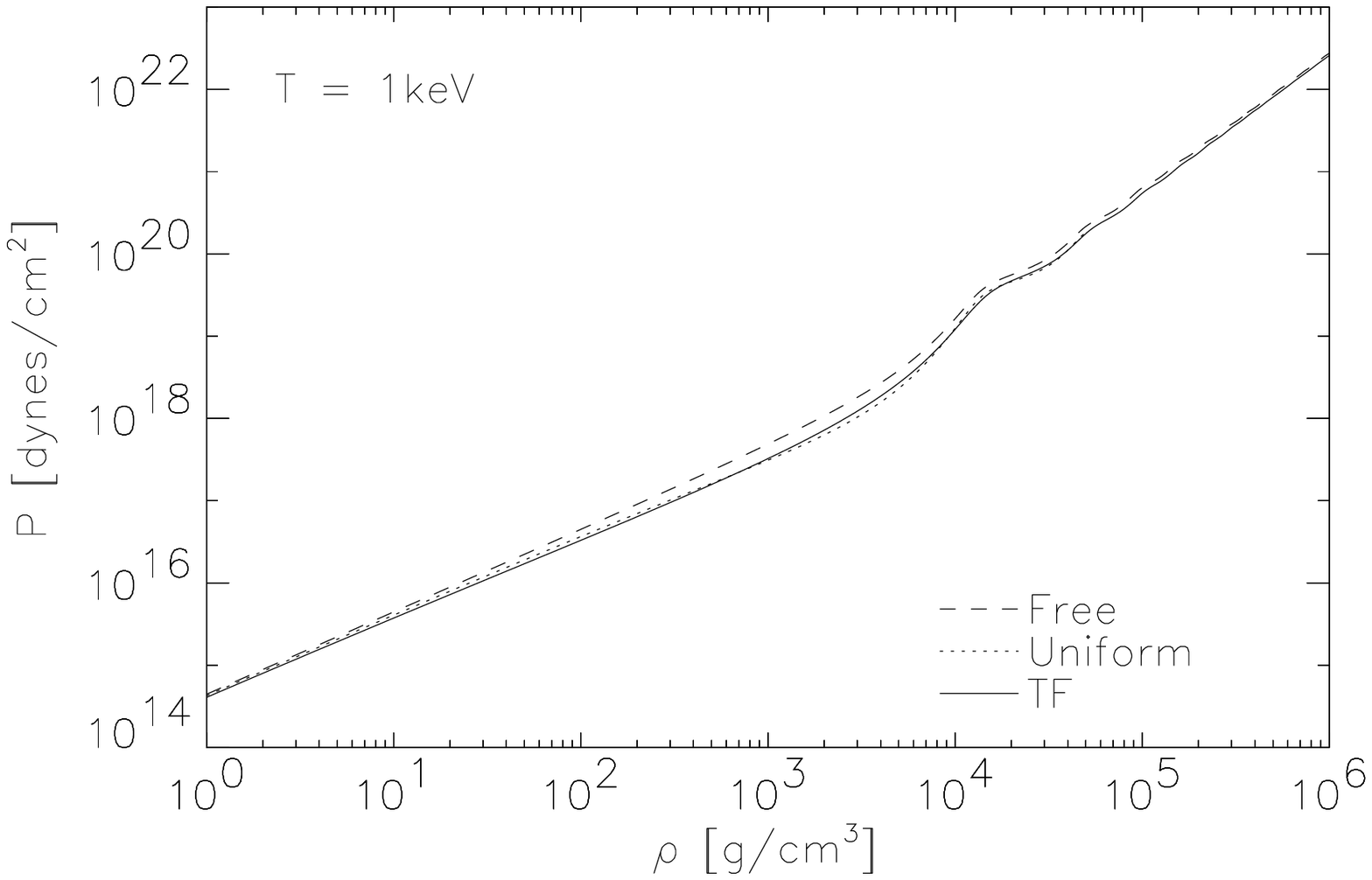}
\caption{}
\label{Fig_PTF-uni-el-b}
\end{figure}

\begin{figure}
\figurenum{7a}
\plotone{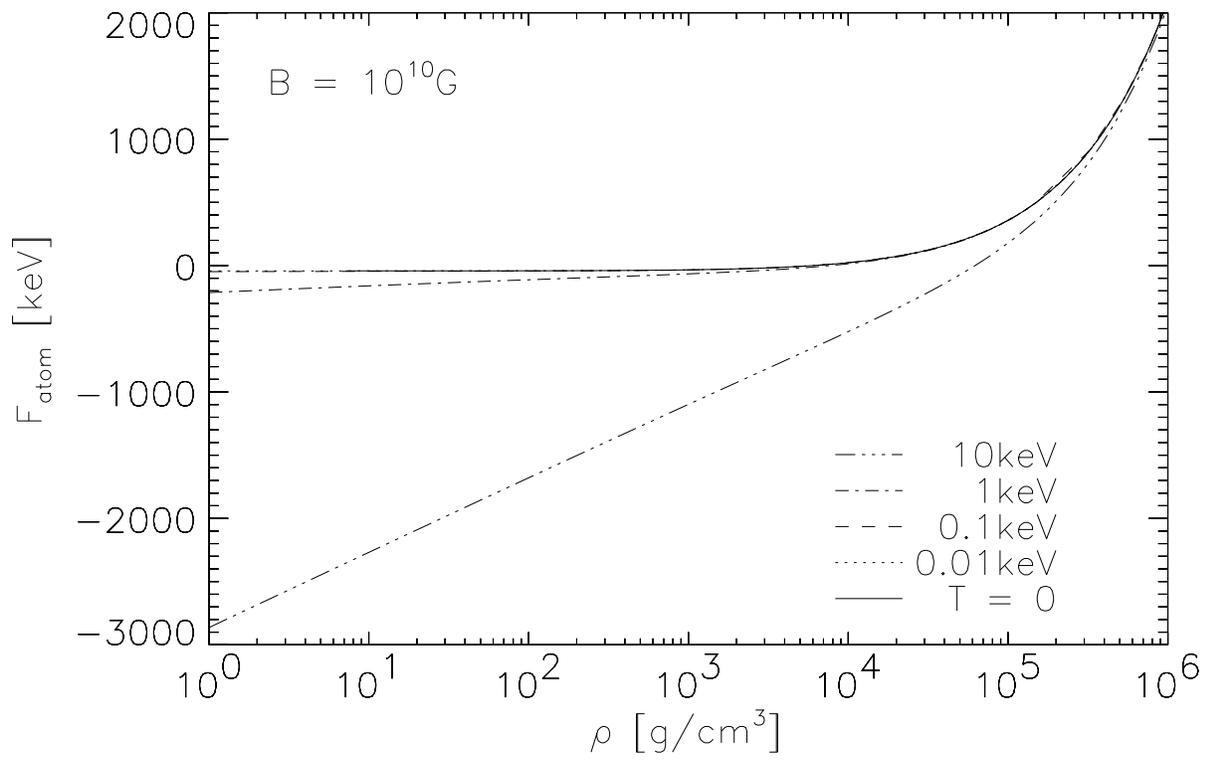}
\caption{
The free energy of iron in bulk, $F_{\rm TF}$, in the Thomas-Fermi 
approximation. Results are shown for several values of the temperature.
(a) $B = 10^{10}$ G. (b) $B = 10^{12}$ G.
\label{Fig_frho}}
\end{figure}


\begin{figure}
\figurenum{7b}
\plotone{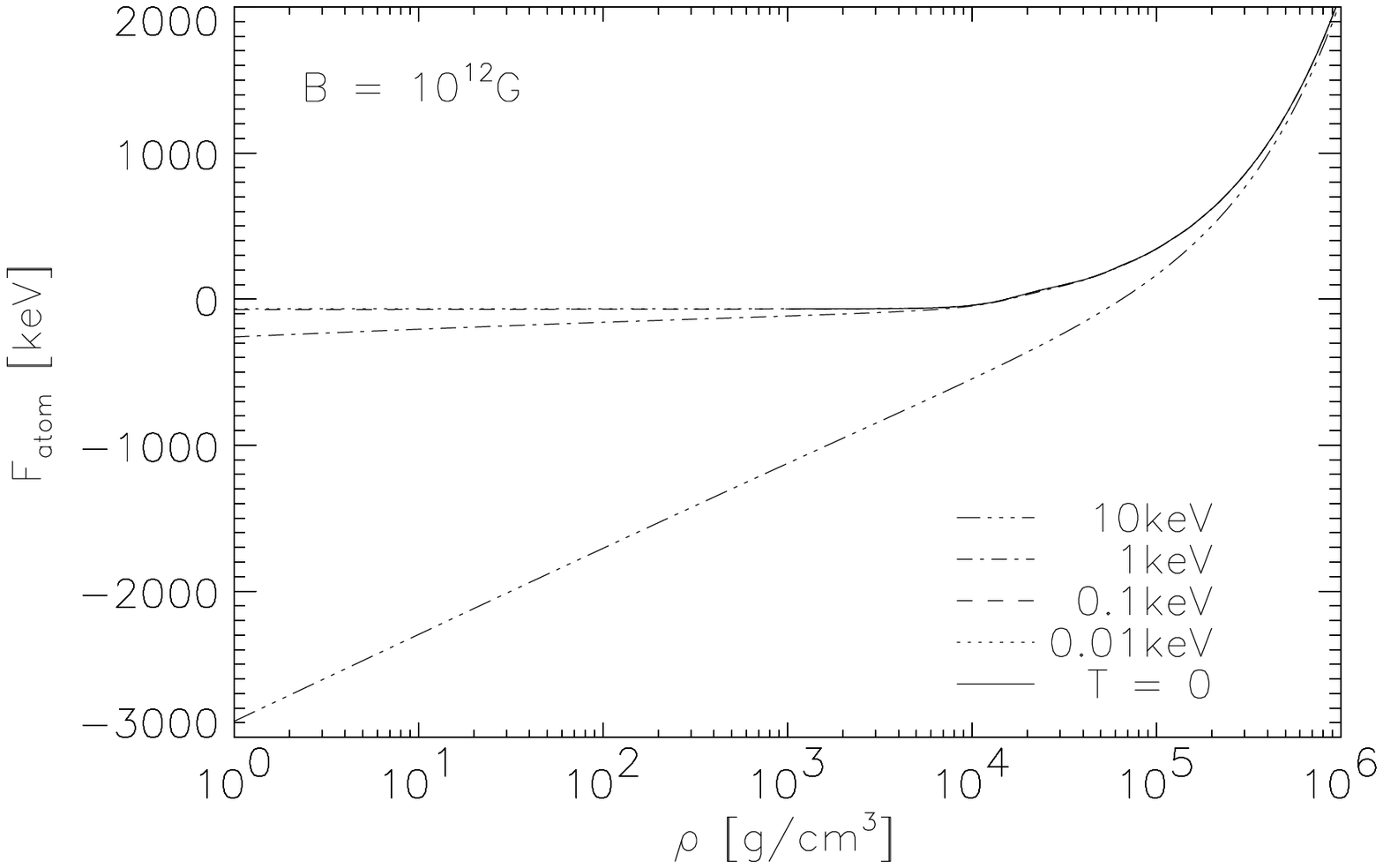}
\caption{}
\label{Fig_frhoB12}
\end{figure}


\begin{figure}
\figurenum{8}
\plotone{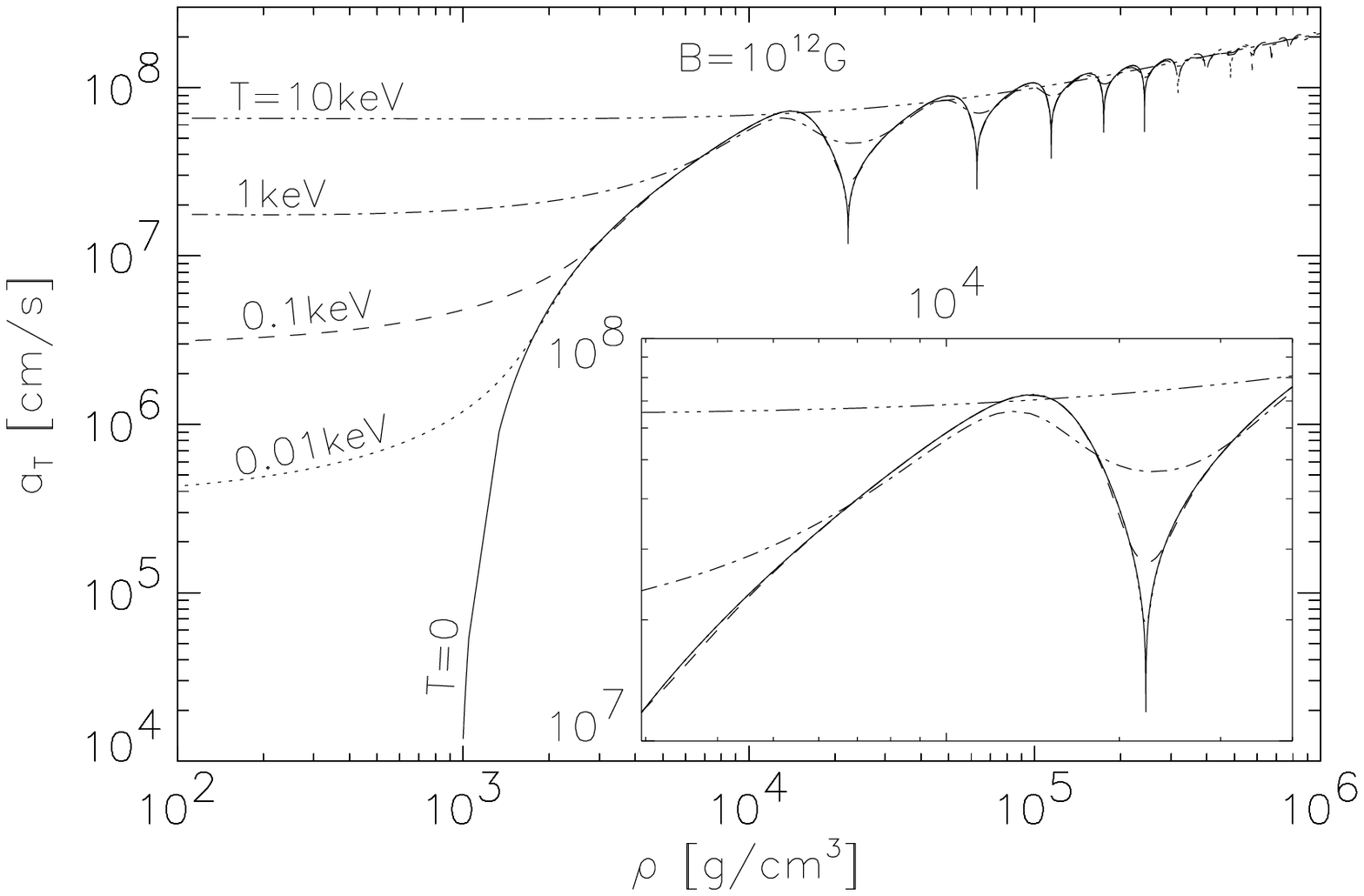}
\caption{
The isothermal sound velocity in bulk matter made of iron in the Thomas-Fermi 
approximation. Each curve corresponds to a given value of the temperature
as indicated. The inset shows details around the peak of the first 
oscillation.
\label{Fig_cs}}
\end{figure}

\begin{figure}
\figurenum{9a}
\plotone{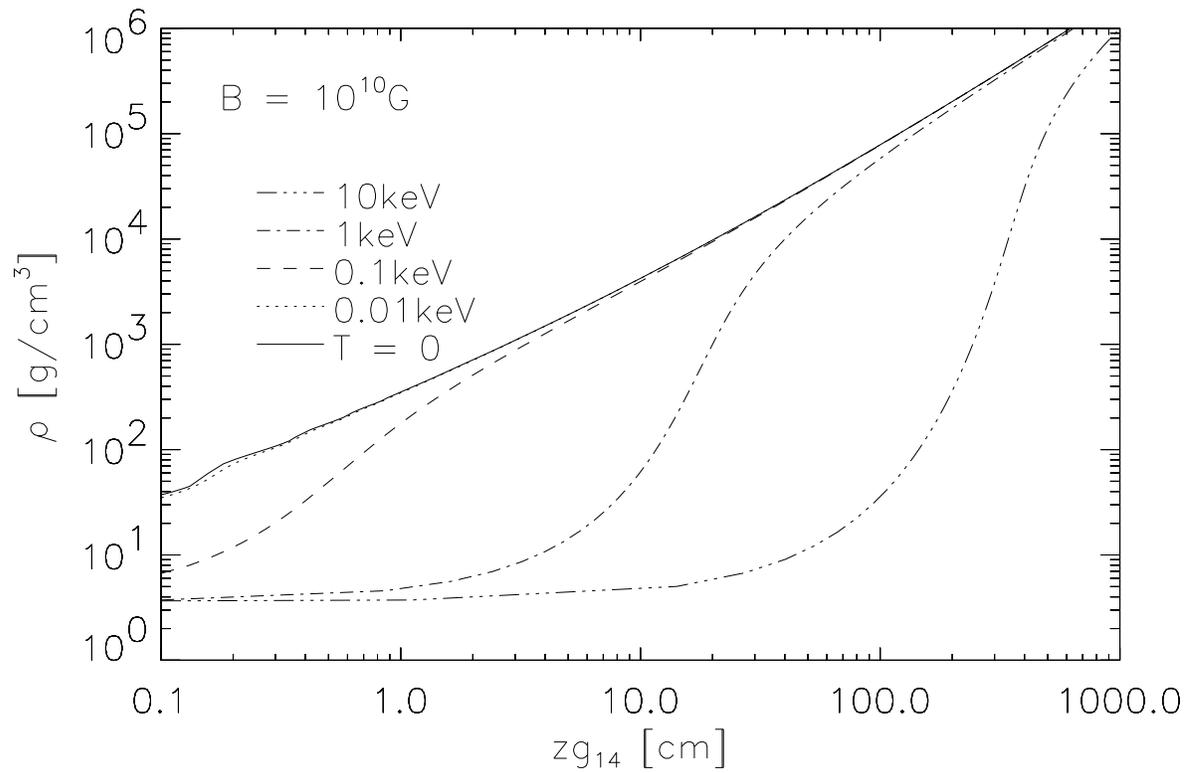}
\caption{
The density of isothermal surface layers of neutron stars as a 
function of depth in the Thomas-Fermi approximation. Results are shown for
five values of the temperature and two values of the magnetic field strength.
The surface gravity is written in units of $10^{14}~ {\rm cm~s}^{-2}$.
Since isothermal atmospheres extend to infinity in the TF approximation the 
depth shown is measured from the point where the surface would be at zero 
temperature. (a) $B = 10^{10}$ G. (b) $B = 10^{12}$ G.
\label{Fig_rhoz}}
\end{figure}

\begin{figure}
\figurenum{9b}
\plotone{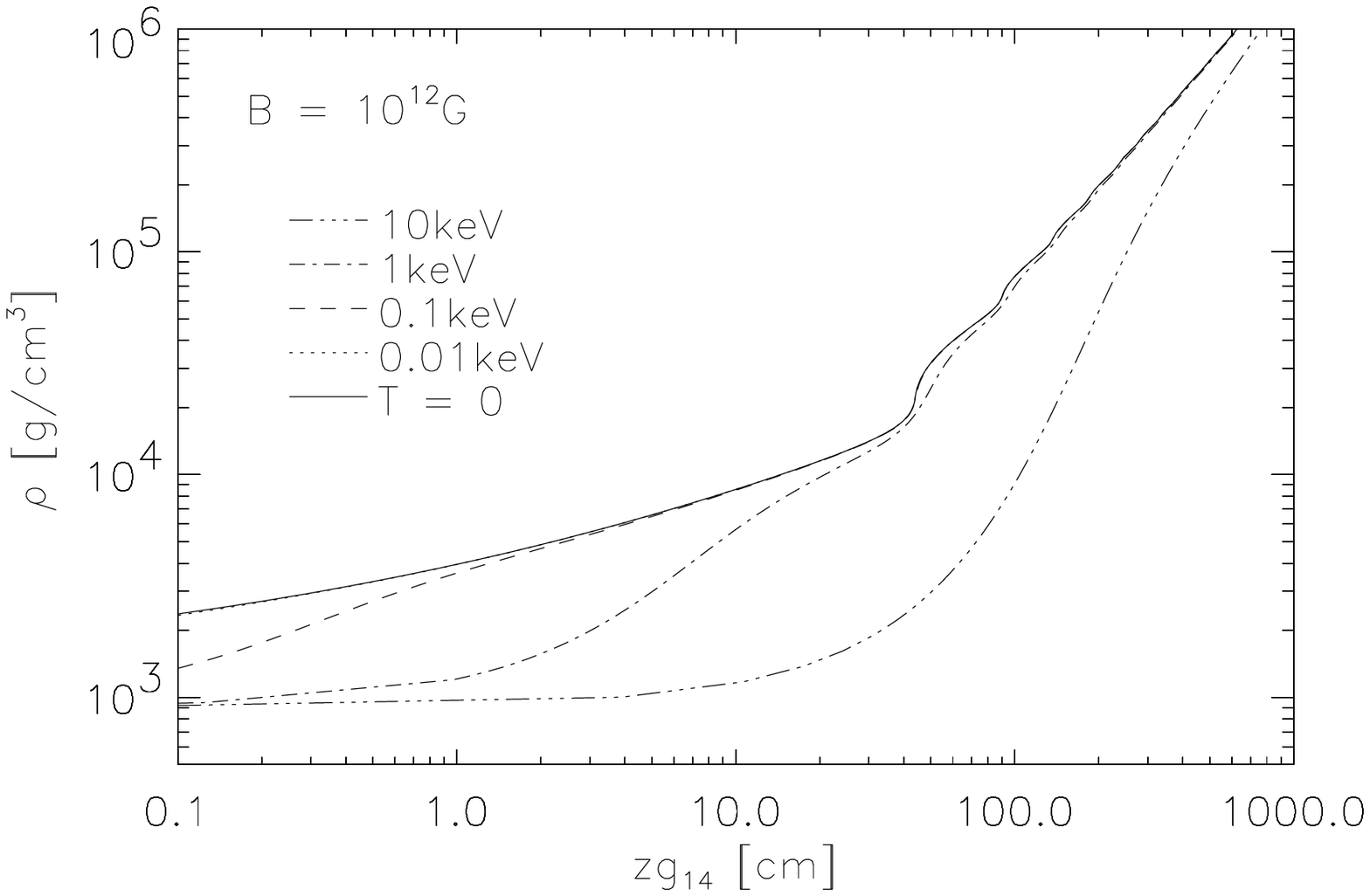}
\caption{}
\label{Fig_rhozB12}
\end{figure}

\begin{figure}
\figurenum{10a}
\plotone{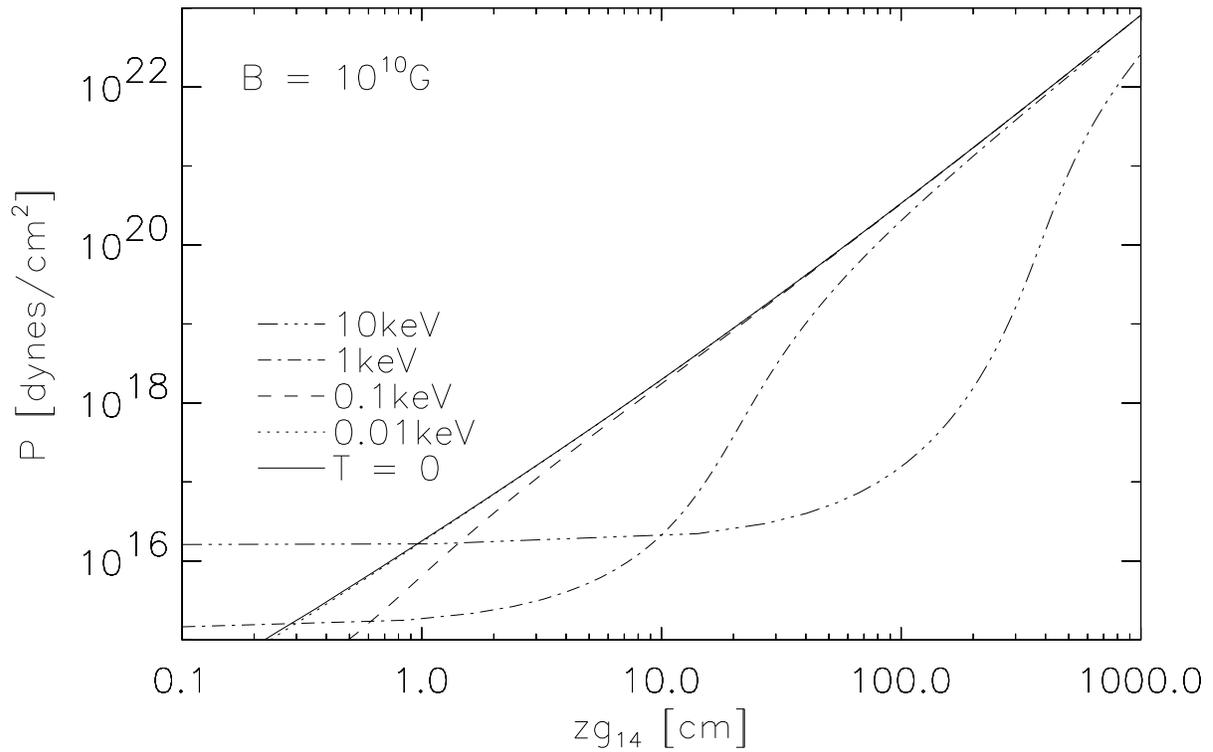}
\caption{
The pressure of isothermal surface layers of neutron stars as a 
function of depth in the Thomas-Fermi approximation. Results are shown for
five values of the temperature and two values of the magnetic field strength.
See the text and the caption for figure \protect{\ref{Fig_rhoz}} for further explanations.
(a) $B = 10^{10}$ G. (b) $B = 10^{12}$ G.
\label{Fig_Pz}}
\end{figure}

\begin{figure}
\figurenum{10b}
\plotone{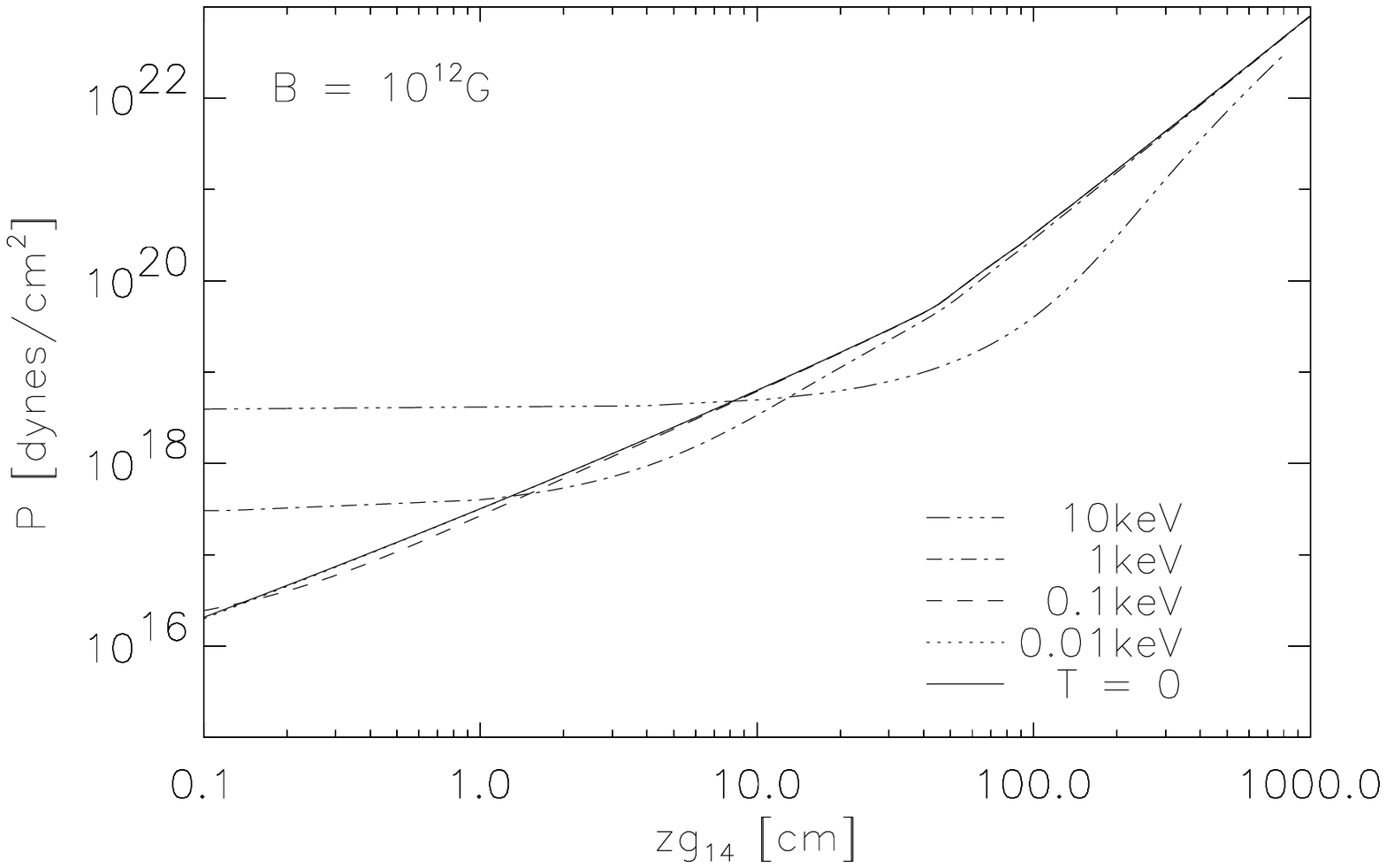}
\caption{}
\label{Fig_PzB12}
\end{figure}



\newpage

\begin{table}[ht]
\begin{center}
  \begin{tabular}{|c|r|r|r|r|r|r|r|}              \hline\hline
$B/T$   &  0.0001  &  0.001  &  0.01  &  0.1  &  1  &  10  &  100  \\ \hline
$n$     &   -      &   -     &  32    &  97   &  97 &  97  &  97   \\ \hline
$P$     &   -      &   50    &  97    &  97   &  97 &  97  &  97   \\ \hline \hline
  \end{tabular}
\end{center}
\caption{Values of $\zeta_{\rm c}$. For $\zeta > \zeta_{\rm c}$ approximations
(\protect{\ref{limit1}}) and (\protect{\ref{limit2}}) are used when
calculating $n$ and $P$, respectively. Where no value is given, the
approximations are used for all values of $\zeta$.
Here, $B$ is measured in units of $10^{12}$ G and $T$ in keV.
\label{Table_zeta}}
\end{table}

\end{document}